\documentclass{article}

\usepackage{PRIMEarxiv}

\usepackage[utf8]{inputenc} 
\usepackage[T1]{fontenc}    
\usepackage{hyperref}       
\usepackage{url}            
\usepackage{booktabs}       
\usepackage{amsfonts}       
\usepackage{nicefrac}       
\usepackage{microtype}      
\usepackage{lipsum}
\usepackage{fancyhdr}       
\usepackage{graphicx}       
\usepackage{amsmath}
\usepackage{natbib}
\usepackage{bm}
\usepackage{rotating}
\usepackage{subfig}

\title{Characterising harmful data sources when constructing multi-fidelity surrogate models
}

\author{
  Nicolau Andr\'{e}s-Thi\'{o} \\
  School of Mathematics and Statistics\\
  The University of Melbourne\\
  ARC Training Centre in\\
  Optimisation Technologies,
  Integrated Methodologies,\\
  and Applications (OPTIMA)\\
  \texttt{nandres@student.unimelb.edu.au} \\
  \And
  Mario Andr\'{e}s Mu\~{n}oz\\
  School of Computer and Information Systems\\
  The University of Melbourne\\
  ARC Training Centre in\\
  Optimisation Technologies,
  Integrated Methodologies,\\
  and Applications (OPTIMA)\\
  \And
  Kate Smith-Miles\\
  School of Mathematics and Statistics\\
  The University of Melbourne\\
  ARC Training Centre in\\
  Optimisation Technologies,
  Integrated Methodologies,\\
  and Applications (OPTIMA)\\
}

\begin{document}
\maketitle

\begin{abstract}
    Surrogate modelling techniques have seen growing attention in recent years when applied to both modelling and optimisation of industrial design problems. These techniques are highly relevant when assessing the performance of a particular design carries a high cost, as the overall cost can be mitigated via the construction of a model to be queried in lieu of the available high-cost source. The construction of these models can sometimes employ other sources of information which are both cheaper and less accurate. The existence of these sources however poses the question of which sources should be used when constructing a model. Recent studies have attempted to characterise harmful data sources to guide practitioners in choosing when to ignore a certain source. These studies have done so in a synthetic setting, characterising sources using a large amount of data that is not available in practice. Some of these studies have also been shown to potentially suffer from bias in the benchmarks used in the analysis. In this study, we present a characterisation of harmful low-fidelity sources using only the limited data available to train a surrogate model. We employ recently developed benchmark filtering techniques to conduct a bias-free assessment, providing objectively varied benchmark suites of different sizes for future research. Analysing one of these benchmark suites with the technique known as Instance Space Analysis, we provide an intuitive visualisation of when a low-fidelity source should be used and use this analysis to provide guidelines that can be used in an applied industrial setting.
\end{abstract}


\section{Introduction}\label{sec1}

In recent years, the development and analysis of techniques for Multi-fidelity Expensive Black-Box (Mf-EBB) problems has gathered a lot of momentum. The growing attention for this type of problem can easily be justified by how often they can be found in industrial design problems such as aerodynamic modelling \citep{peng2023multi} and materials design \citep{khatamsaz2021adaptive}. These \textit{black-box} design problems are characterised by the unknown relationship between design outcomes and decision variables, for which the only way to evaluate decision outcomes is via a deterministic procedure. Each evaluation is deemed to be \textit{expensive}, meaning the amount of sampling of the black-box is severely restricted due to its high computational, monetary or time cost. Furthermore, these \textit{multi-fidelity} problems have multiple sources of information available on the design outcome, with varying degrees of cost and accuracy. The simplest variant of such problems, known as Bi-fidelity Expensive Black-Box (Bf-EBB) problems, have only a single low-fidelity (as well as a high-fidelity) information source available.

The existence of more than one source allows techniques to mitigate the high cost of relying exclusively on a single, very expensive black-box by relying on cheaper (and less accurate) sources of information. This is often achieved via the construction of surrogate models which combine scarce high-fidelity data with more abundant data of lower fidelities. The aim of these models is to accurately predict the outcome value in regions that have not yet been sampled. This information is then used to guide further sampling of the objective function, where the end goal can be to accurately model the relationship of the design variables and the design outcome, or the optimisation of the design problem. A very large proportion of surrogate modelling techniques for Mf-EBB problems are either based on or variations of the seminal work by \citet{kennedy2000predicting}. Their work presents a Bayesian methodology that fuses multiple sources of information into a single surrogate model based on Gaussian processes. Perhaps the most well-known technique derived from this work is the adaptation of Kriging \citep{krige1951statistical, matheron1963principles, jones2001taxonomy}, a surrogate technique that employs only high-fidelity data, to the multi-fidelity setting in the form of the technique known as Co-Kriging \citep{forrester2007multi}. A multitude of studies exist which present variations to this technique, such as different approaches to train the surrogate model \citep{park2018low, shu2019novel}, variations of the surrogate model structure \citep{han2012hierarchical}, different frameworks for integrating multiple non-hierarchical fidelity sources \citep{eweis2022data, cheng2021multi}, and different procedures for guiding further sampling both within the sample space and among different sources \citep{foumani2023multi, yang2022sequential}, among others.

Due to the importance of the accuracy of the surrogate model when conducting design exploration or optimisation, recent studies have analysed the impact that the quality of low-fidelity sources has on the accuracy of a model built from them. \citet{toal2015some} was one of the first to highlight the potential risk in assuming low-fidelity sources can always be relied upon. He focused on Bf-EBB problems where, for a given sample of high and low-fidelity data, he compared the accuracy of a (single-source) Kriging model with the accuracy of a (two-source) Co-Kriging model. He concluded that the correlation between the high- and low-fidelity sources is very important, in fact stating that if the low-fidelity source is not highly correlated with the high-fidelity source, it is recommended to ignore it entirely and train a (single-source) Kriging model instead. Further work in this area \citep{dong2015multi, park2017remarks, liu2018cope, shi2020multi} has reinforced this guideline for low-fidelity data usage. These results have led to the inclusion in new research of an analysis of the test instances employed to assess the performance of new techniques. This analysis is often carried out by measuring the correlation between the high and low-fidelity sources \citep{song2019radial, lv2021multi, toal2023applications}, or by plotting the high-fidelity function values against the low-fidelity function values \citep{zhao2023general, thenon2022sequential} in order to analyse how well the low-fidelity source represents the high-fidelity source.

A recent study by \citet{andres2022bifidelity} however has shown that using a low-fidelity source can be valuable when training a model if it is often locally accurate, even if its overall accuracy is relatively low. More importantly, their work highlighted the potential bias in literature test instances. This bias comes about from the creation procedures of synthetic high- and low-fidelity function pairs, where the quality of the low-fidelity source (either good, average, or bad) is the same throughout the design space. Indeed, it is rare to find synthetic test instances in the literature for which the low-fidelity source is only sometimes locally accurate. To remedy this, a new instance creation procedure was proposed, as well as new measures to assess differences between test instances. This work also showed the lack of and need for an unbiased test suite for Mf-EBB problems, especially when attempting to characterise what constitutes a harmful low-fidelity data source in surrogate model construction. Another key aspect of both the work of \citet{toal2015some} and \citet{andres2022bifidelity} is its synthetic setting. Indeed, whilst the models being assessed are trained with limited data, the characterisation of the high- and low-fidelity sources has so far been conducted using a large amount of data that is not available in practice. The guidelines developed in this setting have enhanced the understanding of the requirements of a useful low-fidelity source and the inner workings of the models being assessed, but these guidelines cannot be directly applied to an industrial setting. A set of guidelines based only on the limited sample available is therefore still lacking in the literature.

The recently developed Instance Space Analysis (ISA) framework \citep{smith2023instance} is well suited to address these literature shortcomings. This method steers away from reporting average algorithm performance across a selected set of test instances. Instead, a visualisation of the space of all possible instances is constructed, allowing an analysis of instance properties (here referred to as \textit{features}) affecting algorithm performance. This analysis provides an objective assessment of bias in a test suite, providing a useful tool to create an unbiased test suite. Furthermore, ISA steers away from the question ``Which algorithm is best?", and instead answers ``Which algorithm is best \textit{for a certain type of instances}?". By using this technique to compare when a single-source surrogate model like Kriging is more accurate than a two-source surrogate model like Co-Kriging, an answer to the question ``When can a low-fidelity source be relied upon?" can be found. Finally, by using features that are calculated only with the available data (and not a much larger set of samples), we can ensure the resulting guidelines can be directly applied to industrial problems.

The work presented in this study can be seen as an extension of the work presented by both \citet{toal2015some} and \citet{andres2022bifidelity}. Its focus is the same as these previous studies, namely the comparison of the accuracy of Kriging and Co-Kriging models in Bf-EBB problems with given high- and low-fidelity data sets. The two main contributions are the construction of an unbiased set of high- and low-fidelity function pairs for future algorithm testing, and the characterisation of harmful low-fidelity data sources, at least when constructing Co-Kriging models, using only the limited data available. This is highly relevant to the literature, as despite many new techniques having been proposed, both Kriging and Co-Kriging are very much still the surrogate models of choice in single-source and two-source EBB problems. The remainder of this paper is structured as follows. Section 2 provides an introduction to ISA and a definition of the instances, features and algorithms used to perform the analysis. Section 3.1 presents the creation of a set of objectively varied function pairs which are then used to generate a set of instances for ISA. Section 3.2 uses these instances to generate the instance space and identify the regions for which Kriging and Co-Kriging should be given precedence. Section 3.3 combines the identification of these regions and the feature value trends to explain why a low-fidelity source is harmful or beneficial. Section 3.4 provides a simplified single-feature analysis in order to provide some easy-to-use guidelines for practitioners in the field. Finally, Section 4 concludes the paper with some closing remarks.

\section{Preliminaries}

\subsection{Instance Space Analysis}

Instance Space Analysis (ISA) \citep{smith2023instance} is based both on the work of \citet{rice1976algorithm}, which proposed using instance features to predict algorithmic performance, and the No-Free Lunch theorems proposed by \citet{wolpert1997no}, the first of which states that ``any two optimization algorithms are equivalent when their performance is averaged across all possible problems". As such, rather than analysing algorithm performance via taking the average performance across a set of test instances, ISA constructs a 2-dimensional instance space to expose similarities and differences among test instances. This instance space is constructed using a chosen set of instance features. By giving a visual representation of where instances are located within the instance space, and how different algorithms perform in different regions of the space, meaningful insights on algorithmic behaviour can be obtained.

\begin{figure*}[!t]
    \centering
    \includegraphics[scale=0.9]{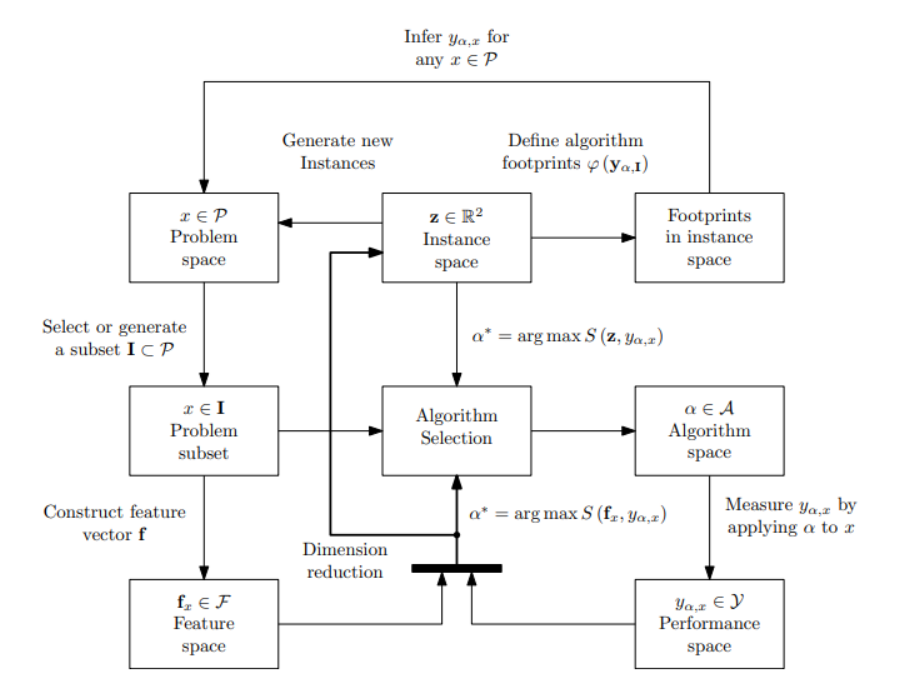}
    \caption{ISA framework \citep{smith2023instance}}
    \label{fig:ISA}
\end{figure*}

Figure \ref{fig:ISA} illustrates the conceptual framework behind ISA. The set $\mathcal{P}$ denotes the \textit{problem space} and contains all possible instances of a particular problem.  The set $\mathbf{I} \subset \mathcal{P}$ is the \textit{subset of instances} for which feature values and algorithmic performance are known. These are the instances that will be used for the analysis of the algorithms. For a given instance $x \in \mathbf{I}$, its feature values are represented by $\mathbf{f}_x \in \mathcal{F}$, with $\mathcal{F}$ representing the \textit{feature space}. The set $\mathcal{A}$ represents the \textit{algorithm space}, i.e. the set of all algorithms available to solve the chosen problem. The performance of a given algorithm $\alpha \in \mathcal{A}$ in an instance $x \in \mathbf{I}$ is given by $y_{\alpha,x} \in \mathcal{Y}$, where $\mathcal{Y}$ represents the \textit{performance space}. A user-defined measure of ``good" performance, either absolute or relative, is required to compare algorithms and analyse in which regions the performance of a particular algorithm is considered to be good. Finally, ISA aims to learn the mapping $S(\cdot)$ from a position of an instance in the 2-dimensional instance space to the algorithm with the best performance. This mapping is used to estimate the \textit{footprint} $\varphi(\mathbf{y}_{\alpha, \mathbf{I}})$ of each algorithm. This is defined as a region within the instance space for which ISA has statistically inferred good performance. This visual property of algorithm performance is what can be used to draw conclusions on the strengths and weaknesses of algorithms, and for which types of instances a particular algorithm should be used. A MATLAB toolkit that performs ISA automatically \citep{isatoolkit} is publicly available, and the online tool Melbourne Algorithm Test Instance Library with Data Analytics (MATILDA) \citep{matilda} provides tutorials, examples and an interface for online analysis.

The collection $\{\mathcal{P}, \mathcal{F}, \mathcal{A}, \mathcal{Y}\}$ denotes the problem's \textit{metadata}. In practice, before running ISA one must choose subsets from each of the elements in the metadata. That is, one must choose which instances, features, algorithms and algorithm performance measures to use in the analysis. Choosing both the algorithms and the performance is conditional on the user's need, namely what kind of analysis is required. Furthermore, a large set of features can initially be used, as the implemented toolkit provides an automatic selection procedure that only retains the most relevant features when creating the instance space. Choosing a subset $\mathbf{I} \subset \mathcal{P}$ however can be a harder problem, as recently shown by \citet{alipour2023enhanced}. Their work showed that over-representation of types of instances in $\mathbf{I}$ can lead to bias in the overall analysis of ISA. They proposed an instance filtering technique that removes ``similar" instances to generate an unbiased test suite. Two instances $x_i$ and $x_j$ with feature vectors $\mathbf{f}_i$ and $\mathbf{f}_j$ are deemed to be similar if $\|\mathbf{f}_i - \mathbf{f}_j\| \leq \theta$, where $\theta$ is a user-defined similarity threshold. The instance filtering algorithm simply iteratively removes instances from an initial set if at least one other ``similar" instance is still in the set. This generates a set of ``dissimilar" instances, denoted $\mathcal{D}_{\theta}$, so that for every pair of instances $x_i,x_j \in \mathcal{D}_{\theta}$, $\|\mathbf{f}_i - \mathbf{f}_j\| > \theta$.

Based on both the algorithm performance of each algorithm, and a user-defined measure of ``good" performance, a binary performance vector $\boldsymbol{\delta}_i$ for each instance $i$ can be defined, where the $k^{th}$ entry of $\boldsymbol{\delta}_i$ is 1 if the performance of algorithm $\alpha_k$ is deemed ``good" for instance $i$, and 0 otherwise. As the algorithm performance for instances that are deemed to be similar is assumed to be the same, a more intricate instance filtering procedure uses both the feature vectors $\mathbf{f}_i$ and binary performance vectors $\bm{\delta}_i$ when choosing which instances to filter out. Specifically, starting from a set of instances $\mathbf{I}$, an instance $i$ is removed from $\mathbf{I}$ if there exists an instance $j$ with $\|\mathbf{f}_i - \mathbf{f}_j\| \leq \theta$ and $\bm{\delta}_i = \bm{\delta}_j$. Instances that are similar feature-wise but have different binary performances are said to violate the similarity assumption. These instances should be kept in order to maintain valuable information in the set. Therefore, in this second filtering procedure, a set of dissimilar instances $\mathcal{D}_{\theta}$ is found as well as the set of instances $\mathcal{V}_{\theta}$ which violate the similarity assumption. The chosen test instances lie in the union of these two sets, known as the critical set $\mathcal{C}_{\theta} := \mathcal{D}_{\theta} \cup \mathcal{V}_{\theta}$. This set satisfies the property that for every pair of instances $x_i,x_j \in \mathcal{C}_{\theta}$ either $\|\mathbf{f}_i - \mathbf{f}_j\| > \theta$ or $\bm{\delta}_i \neq \bm{\delta}_j$. The user-defined $\theta$ determines the trade-off between the diversity of the final set of instances, and uniformity in feature-distance between instances. In order to choose this $\theta$, a uniformity measure $\mathfrak{u}_{D_{\theta}}$ is introduced, with

\begin{align*}
    \mathfrak{u}_{\mathcal{D}_\theta} = 1- \frac{\sigma_{\mathcal{D}_{\theta}^{NN}}}{\mu_{\mathcal{D}_{\theta}^{NN}}}\\
\end{align*}

\noindent where $\mathcal{D}_{\theta}^{NN} = \{\min_{j \in \mathcal{D}_{\theta}, j \neq i}\|\mathbf{f}_i - \mathbf{f}_j\| \quad | \quad i \in \mathcal{D}_\theta\}$ is the nearest neighbour feature distance for the instances in $\mathcal{D}_\theta$, and $\mu_{\mathcal{D}_{\theta}^{NN}}$ and $\sigma_{\mathcal{D}_{\theta}^{NN}}$ are its mean and variance, respectively. The authors recommend scaling $ \mathfrak{u}_{\mathcal{D}_\theta}$ so that its values lie between 0 and 1, and choosing the smallest $\theta$ for which $ \mathfrak{u}_{\mathcal{D}_\theta} \geq 0.5$. Finally, it is worth noting that it is possible to implement the filtering procedure so that certain instances are prioritised when choosing which instances to keep. This characteristic will be employed below to retain as many literature instances as possible. For further details on this procedure, the reader is directed to the work of \citet{alipour2023enhanced}.

\subsection{Instances, Features and Algorithms}

\subsubsection{Instances}

The predominant characteristic of Bf-EBB problems is the existence of two black boxes that can be queried for information. Both the high-fidelity black box, denoted $f_h$, and the low-fidelity black box, denoted $f_l$, are defined within a hypercube $\Omega$. That is

\begin{align*}
    f_h,f_l: \quad &\Omega \rightarrow \mathbb{R}\\
   \Omega = [x^\bot_1, x^\top_1] &\times \dots \times [x^\bot_d, x^\top_d]\\
\end{align*}

\noindent where $\mathbf{x}^\bot = (x^\bot_1, \dots, x^\bot_d)$ and $\mathbf{x}^\top = (x^\top_1, \dots, x^\top_d)$ are the vectors representing the lower and upper bounds of $\Omega$, and $d \in \mathbb{N}$ is the dimension (i.e. number of variables) of the problem. It is worth noting that in some cases in the literature, one of or both sources can be assumed to be stochastic. That is not the case in this study, as we assume that both $f_h$ and $f_l$ are deterministic. Whilst $f_h$ is assumed to be expensive, how this cost is translated to the problem statement can vary. The two most common settings are providing an existing sample of both $f_h$ and $f_l$, or providing a maximum number of times that an algorithm can query $f_h$. Furthermore, as discussed in the introduction, the aim can be to either minimise $f_h$, or to train a surrogate of $f_h$ which is as accurate as possible.

In this study, the focus is put on what is arguably the simplest variant; where the sampling of both $f_h$ and $f_l$ has already been conducted, no further sampling is possible, and the only decision is to choose which surrogate model to train. An instance in this study is therefore defined by the tuple $(f_h,f_l,n_h,n_l)$, where $n_h$ and $n_l$ are the number of locations at which the values of $f_h$ and $f_l$ are known, respectively. As both $f_h$ and, to a lesser extent, $f_l$ are considered to be expensive, $n_h$ and $n_l$ should be kept relatively small. Therefore the values used are $n_l \in \{4d, 8d, 12d, 16d, 20d\}$ and $n_h \in \{2d, 4d, \dots, 18d, 20d\}$, with $n_h \leq n_l$. In order to generate the set $\mathbf{I}$ however, a choice must be made of which function pairs pairs $(f_h,f_l)$ to include from a large candidate pool. This is expanded on in the next section.

\subsubsection{Algorithms and Performance}

The aim of this study is to characterise when a low-fidelity source can be harmful when constructing a surrogate model. As such, two types of models are compared, namely a surrogate model which is trained using only high-fidelity data, and a surrogate model which is trained using both high and low-fidelity data. As the findings should be as widely applicable as possible to the field, two well-established models are used, namely Kriging \citep{krige1951statistical, matheron1963principles, jones2001taxonomy} and Co-Kriging \citep{forrester2007multi}. For an in-depth mathematical derivation, the reader is directed to the work of \citet{jones2001taxonomy} and \citet{forrester2007multi}; an introduction to both is also given in the implementation of both techniques used in this work \citep{nandres2022methods}. The wide usage of these two techniques in the literature can be attributed to three main components. The first is their strong theoretical backing, which assumes the observed data from either source is the result of a multivariate random normal variable, and as such the training of the model's hyperparameters is done by finding the maximum likelihood estimators of a probability distribution. The second is the high accuracy these models tend to have even with relatively small amounts of data. This is in part due to the high number of hyperparameters which allows the models to capture how the objective function changes along each of the input dimensions. It is worth noting that the training of these hyperparameters can take a long time when a lot of data is available. This however is not a problem for Bf-EBB problems, as one of its key assumptions is that the data available is scarce. Finally, an attractive characteristic of these models once they have been trained is the existence of not only a prediction of the objective function value, but also an error estimate for the prediction. Whilst this characteristic is of no benefit if no further sampling is allowed, further sampling can be balanced between exploration and exploitation of the sample space through the use of this error metric.

By taking $\mathcal{A} = \{$Kriging, Co-Kriging$\}$, the next requirement to perform ISA is the definition of model performance, as well as a definition for ``good" model performance. The performance of the two methods in an instance $(f_h,f_l,n_h,n_l)$ is assessed through a statistical comparison via repetitions of the training of each of the models. At each repetition, the sets  $\mathbf{X}_h, \mathbf{y}_h, \mathbf{X}_l$ and $\mathbf{y}_l$ defined as\\

\begin{align*}
    \mathbf{X}_l &= \{\mathbf{x}^l_{1}, \mathbf{x}^l_{2}, \dots, \mathbf{x}^l_{n_l}\} \subset \Omega\\
    \mathbf{X}_h &= \{\mathbf{x}^h_{1}, \mathbf{x}^h_{2}, \dots, \mathbf{x}^h_{n_h}\}\subset \Omega\\[1ex]
    \mathbf{y}_l &= \{f_l(\mathbf{x}_1^l), \dots, f_l(\mathbf{x}_{n_l}^l)\}\\
    \mathbf{y}_h &= \{f_h(\mathbf{x}_1^h), \dots, f_h(\mathbf{x}_{n_h}^h)\}\\
\end{align*}

\noindent are constructed, with $|\mathbf{X}_l| = n_l$ and $|\mathbf{X}_h| = n_h$. The set $\mathbf{X}_l$ is a sampling plan of $f_l$. It is obtained by creating a random Latin Hypercube Sampling (LHS) plan of $\Omega$. It is possible however for LHS plans to not be well spread out. The initial random LHS is therefore optimised to be locally optimal in terms of the minimum distance between any pair of points. This is achieved by iteratively swapping one of the entries between two pairs of points, as long as this increases the distance between the two closest points in the set. Once this can no longer be achieved, the resulting set is locally optimal. Similarly, the set $\mathbf{X}_h$ is constructed as a subset of $\mathbf{X}_l$ and also made locally optimal in terms of the minimum distance. This is achieved by starting with a random subset of $\mathbf{X}_l$, and iteratively swapping a point inside the subset for a point outside the subset as long as this increases the minimum distance between any pair of points. Once this can no longer be done, the resulting set is locally optimal. It is worth noting that both of these ``brute force" approaches can be lengthy for larger sample sizes. For this reason as well as for reproducibility purposes, the sampling plans used in this study are made available in the implementation repository \citep{nandres2022methods}.

Once the sets $\mathbf{X}_h, \mathbf{y}_h, \mathbf{X}_l$ and $\mathbf{y}_l$ have been generated, a Kriging model is trained using $\mathbf{X}_h$ and $\mathbf{y}_h$, and a Co-Kriging model is trained using $\mathbf{X}_h, \mathbf{y}_h, \mathbf{X}_l$ and $\mathbf{y}_l$. Note that in the applied setting the accuracy of a surrogate model cannot be assessed as it cannot be compared with the values of the true objective function. However as is standard practice when using synthetic test functions, here a large sample set is used to assess model accuracy. That is we generate the datasets $\mathbf{X}$ and $\mathbf{y}$ defined as

\begin{align*}
    \mathbf{X} &= \{\mathbf{x}_{1}, \mathbf{x}_{2}, \dots, \mathbf{x}_{N}\} \subset \Omega\\
    \mathbf{y} &= \{f_h(\mathbf{x}_1), \dots, f_h(\mathbf{x}_{N})\}\\
\end{align*}

\noindent with $N = 1000d$. Given a surrogate model $s: \Omega \rightarrow \mathbb{R}$, its accuracy is assessed using its Pearson's sample correlation $P_{corr}$ with the objective function, given by

\begin{align*}
    P_{corr} = \frac{1}{N-1}&\left(\frac{\sum_{i = 1}^{N} (f_h(\mathbf{x}_i) - \bar{y})(s(\mathbf{x}_i) - \bar{S})}{s_{Y} s_{S}}\right)\\
    \text{where } \qquad \bar{y} &= \frac{1}{N}\sum_{i = 1}^{N}f_h(\mathbf{x}_i)\\
    s_{Y} &= \left[\frac{\sum_{i = 1}^{N}(f_h(\mathbf{x}_i) - \bar{y})^2}{N-1}\right]^{1/2}\\
    \bar{S} &= \frac{1}{N}\sum_{i = 1}^{N}s(\mathbf{x}_i)\\
    s_{S} &= \left[\frac{\sum_{i = 1}^{N}(s(\mathbf{x}_i) - \bar{S})^2}{N-1}\right]^{1/2}\\
\end{align*}

Here, a high $P_{corr}$ indicates high model accuracy as the model predicts the behaviour of the objective function. Finally, as the analysis of interest is not to find when Kriging and Co-Kriging perform well, but when they perform well comparatively, the performance measure used is relative in nature. That is, after 40 repetitions of generating the datasets and training the models, the resulting 40 Kriging and Co-Kriging model accuracies are compared using the Wilcoxon test \citep{wilcoxon1992individual}. The performance value of Kriging on a particular instance is the $p$-value of a single-tailed Wilcoxon test with the null hypothesis that the correlation of the Kriging model with $f_h$ is statistically no less than 0.001 than the correlation of the Co-Kriging model with $f_h$. In essence, the $p$-value of this test gives the probability that the Kriging model accuracy is either higher or almost identical (within a tolerance of 0.001) to the accuracy of the Co-Kriging model. If this $p$-value is larger than 0.5, the performance of the Kriging model is assumed to be good. An identical procedure is followed to obtain the $p$-value of the Co-Kriging model. It is worth noting that both $p$-values can be large (i.e. the performance is ``good" for both) if the performance of both models is statistically similar.





\subsubsection{Features}

The features used in this study characterise the relationship between $f_h$ and $f_l$, the landscape of $f_h$, $f_l$, and $f_h - f_l$ (the difference between the two sources), as well as the amount of data available when training the models. The relationship features used are the Correlation Coefficient ($CC$) and Relative Root Mean Squared error ($RRMSE$) proposed by \citet{toal2015some} which indicate the overall quality of $f_l$ relative to $f_h$, as well as the Local Correlation Coefficient features $LCC^r_{mean}, LCC^r_{sd}, LCC^r_{coeff}$ and $LCC^{r}_p$ proposed by \citet{andres2022bifidelity} which quantify the distribution of the local quality of $f_l$. The value of $p$ gives a threshold of ``good" local correlation, and the value of $r$ impacts the radius of the $d$-dimensional ball used to calculate the local correlations. Here $p \in \{0.1, 0.2, \dots, 0.9, 0.95, 0.975\}$, and $r = \{0.2, 0.2^{1/d}\}$, where $d$ is the problem dimension. The value $r = 0.2$ is used as it has been shown to lead to helpful features in a synthetic setting, and the value $r = 0.2^{1/d}$ is used here for the first time to assess whether growing the hypervolume of the sphere constant relative to the volume of a hypersphere encompassing the whole sample space also leads to good features. The mathematical definition of these features in given in Appendix \ref{secA1}. The landscape features are taken from the R package flacco \citep{KerschkeT2019flacco}, which is made up of 17 feature sets, such as classic Exploratory Landscape Analysis (ELA) or Information Content feature sets. Not all features in the package are used however; features that split up the sample space into cells cannot be used on problems of high dimension ($d \geq 20$), and features that are related to the function's domain and range are not used as they are not comparable between functions. The features which characterise the amount of data available are the actual sizes of the data i.e. $n_h$ and $n_l$, the sizes relative to the dimension $\frac{n_h}{d}$ and $\frac{n_l}{d}$, and the ratio of high to low-fidelity data $\frac{n_h}{n_l}$. Finally, the problem dimension itself is taken as a feature as well. All of the features used are given in Table \ref{tab:features}, along with their range of values.


\renewcommand{\arraystretch}{1.3}
\begin{sidewaystable*}
\tiny
    \centering
    \begin{tabular}{l c p{0.6\textwidth} c}
        \toprule
        Method & Feature name & Description & Range\\ \midrule
        Overall quality of $f_l$\citep{toal2015some} & $CC$ & Overall correlation between $f_h$ and $f_l$ & $[0,1]$\\
        & $RRMSE$ & Overall error between $f_h$ and $f_l$ & $[0,\infty)$\\
        Local quality of$f_l$\citep{andres2022bifidelity} & $LCC^{r}_{p}$ & Proportion of time the local correlation between $f_h$ and $f_l$ is above a given $p \in [0,1]$ & [0,1]\\
        & $LCC^{0.2}_{mean}$ & Average local correlation between $f_h$ and $f_l$ & [0,1]\\
        & $LCC^{0.2}_{sd}$ & Standard deviation of the local correlation between $f_h$ and $f_l$ & $[0,1]$\\
        & $LCC^{0.2}_{coeff}$ & Coefficient of variation of the local correlation between $f_h$ and $f_l$ & $[0,\infty)$\\
        Distribution (ELA distr) \citep{mersmann2011exploratory} & Skewness & Skewness of the objective function values & $\mathbb{R}$\\
        & Kurtosis & Kurtosis of the objective function values & $\mathbb{R}$\\
        & Peaks & Estimated number of peaks in the objective function & $[1,\infty)$\\
        Levelset prediction (ELA levelset) \citep{mersmann2011exploratory} & $MMCE_{lda}^q$ & Mean misclassification error of a predictive linear model with data quantile split $q \in \{0.1, 0.25, 0.5\}$ & $[0,1]$\\
        Meta modeling (ELA meta)\citep{mersmann2011exploratory} & $\bar{R}^2_{L}$ & Adjusted $R^2$ of a linear model without interactions & $[0,1]$\\
        & $\bar{R}^2_{LI}$ & Adjusted $R^2$ of a linear model with interactions & $[0,1]$\\
        & $\bar{R}^2_{Q}$ & Adjusted $R^2$ of a quadratic model without interactions & $[0,1]$\\
        & $\bar{R}^2_{QI}$ & Adjusted $R^2$ of a quadratic model with interactions & $[0,1]$\\
        & $CN_{L}$ & Ratio of the minimum and maximum absolute coefficients of a linear model without interactions & $[0,\infty)$\\
        & $CN_{Q}$ & Ratio of the minimum and maximum absolute coefficients of a quadratic model without interactions & $[0,\infty)$\\
        Information content\citep{munoz2014exploratory} & $H_{max}$ & Maximum information content of fitness sequence & $[0,1]$\\
        & $\epsilon_{S}$ & Settling sensitivity; epsilon for which sequence contains 0 almost exclusively & $\mathbb{R}$\\
        & $\epsilon_{max}$ & Value for which $H(\epsilon_{max}) = H_{max}$ & $\mathbb{R}$\\
        & $\epsilon_{ratio}$ & Ratio of partial information sensitivity & $\mathbb{R}$\\
        & $M_0$ & Initial partial information & $[0,1]$\\
        Nearest Better Clustering \citep{preuss2012improved} & $NBC_{mean}$ & Ratio of the mean distance of a point's closest neighbour, and a point's better closest neighbour & $[0,1]$\\
        & $NBC_{sd}$ & Ratio of the standard deviation of the distance of a point's closest neighbour, and a point's better closest neighbour& $[0,1]$\\
        & $NBC_{coeff}$ & Coefficient of variation of the ratios of the distance of the closest neighbour, and the distance of the better closest neighbour& $[0,\infty)$\\
        & $NBC_{corr}$ & Correlation between the distance of a point's nearest neighbour and a point's nearest neighbour with a lower objective function & $[-1,1]$\\
        & $NBC_{fitcorr}$ & Correlation between a point's objective function value and its ``in-degree" & $[-1,1]$\\
        Principal component analysis & $PCA_{corr}$ & Relative amount of principal components required to explain high variability in the sample correlation\\
        & $PCA_{cov}$ & Relative amount of principal components required to explain high variability in the sample covariance& $[0,1]$\\
        Dispersion \citep{lunacek2006dispersion} & $\overline{DISP}_{\epsilon}$ & Ratio of the mean distance between all points, and the mean distance between the $\epsilon\%$ best points, $\epsilon \in \{2, 5, 10\}$ & $[0, \infty)$\\
        & $\widetilde{DISP}_{\epsilon}$ & Ratio of the median distance between all points, and the median distance between the $\epsilon\%$ best points, $\epsilon \in \{2, 5, 10\}$ & $[0, \infty)$\\
        Data budget & $B_h = n_h$ &  Number of high-fidelity samples available & $[2,400]$\\
        & $B_l = n_l$ &  Number of low-fidelity samples available & $[2,400]$\\
        & $B^r_h = n_h/d$ &  Relative number of high-fidelity samples available & $[2,20]$\\
        & $B^r_l = n_l/d$ &  Relative number of low-fidelity samples available & $[2,20]$\\
        & $B^r = n_h/nl$ &  Ratio of number of high-fidelity and low-fidelity samples & $[0,1]$\\
        Problem dimension & $d$ & & $[1,20]$ \\
        \bottomrule
    \end{tabular}
    \caption{A short description of the features used in this study, as well as the range of feature values for each feature.}
    \label{tab:features}
\end{sidewaystable*}

The feature values must be processed before being used both for instance filtering and ISA. It is a desirable property for the processed feature values to have similar ranges. Therefore, for the features which are known to be bounded a linear transformation is applied so that the range of the transformed features is $[-2,2]$. This linear transformation is also applied to the dispersion features $\overline{DISP}_{\epsilon}$ and $\widetilde{DISP}_{\epsilon}$, as despite being unbounded in theory, in practice the distribution of the feature values is close to uniform $[0,1]$. The Box-Cox transformation is applied to the remaining (unbound) features to remove the effect of outliers, and then transformed a second time in order to produce feature values that have a standard normal distribution. Finally, any remaining outliers are bound to the range $[-4,4]$. As $95\%$ of the values are within $[-2,2]$, this makes them comparable to the features processed with a linear transformation.

When choosing a set of function pairs $(f_h,f_l)$ (as described in the next section) in order to construct the benchmark suites, the feature values are calculated using a very large sample of size $1000d$, where $d$ is the problem dimension. Once the set of instances $\mathbf{I}$ has been constructed, all features are calculated using only the sample used to train the Kriging and Co-Kriging models. As for each instance, 40 repetitions are conducted (and therefore 40 high- and low-fidelity data sets are generated), and the sample feature values are the average of the feature value calculated with each of the 40 data sets.

\section{ISA of Bf-EBB surrogate modelling with fixed sample}

\subsection{Generating an unbiased test suite}

As stated in the previous Section, choosing a set of function pairs $(f_h,f_l)$ to generate the set of instances $\mathbf{I}$ is not a trivial problem. A large set of candidates has been implemented to be considered for this purpose. This set of candidates consists of over 200 literature function pairs, supplemented by over 80,000 function pairs generated by the instance-generating procedure proposed by \citet{andres2022bifidelity}. These disturbance-based function pairs are generated by starting with a chosen high-fidelity function, and adding a ``disturbance" either around a particular objective function value, or near disturbance ``centres" in the sample space. The literature function pairs are gathered from a variety of studies \citep{rajnarayan2008multifidelity,march2012provably, liu2016multi, liu2018cope, wu2020active, shi2020multi, dong2015multi, xiong2013sequential, simulationlib, park2017remarks} and include function pairs of dimensions $d \in \{1, 2, 3, 4, 5, 6, 8, 10, 20\}$. The high-fidelity functions from this set as well as functions from the COCO test suite \citep{hansen2020cocoplat} with $d \in \{2, 4, 6, 8, 10\}$ are used to generate the disturbance-based function pairs. For further details on the implemented function pairs, the reader is directed to the publicly available implementation \citep{nandres2022benchmarks}.

As algorithm performance cannot be assessed for such a large candidate set of function pairs, the simpler instance filtering procedure is first used which relies only on feature values. The aim here is to generate a set of function pairs that can be used in future research to allow for comparable analysis across studies. In order to prevent a certain type of feature to have an unwanted weight when deciding which instances to filter out, only the ``real" (i.e. calculated with a large sample) features $CC$, $RRMSE$, $LCC^{0.2^{1/d}}_{0.5}$, $LCC^{0.2^{1/d}}_{0.95}$, $LCC^{0.2^{1/d}}_{sd}$, Skewness, Kurtosis, Peaks, $MMCE^{0.25}_{lda}$, $\bar{R}^2_{Q}$, $CN_{Q}$, $H_{max}$, $H_0$, $\epsilon_{max}$, $\epsilon_{ratio}$, $\overline{DISP}_{10}$, $NBC_{mean}$, $NBC_{sd}$ and $PCA_{cov}$ are used, calculated for $f_h$, $f_l$ and $f_h - f_l$. The filtering prioritises literature function pairs, only choosing newly generated disturbance-based function pairs if they are sufficiently different from every literature function implemented. The filtering leads to a set of 312 function pairs, 42 from the literature and 280 disturbance-based function pairs. It is worth noting here that it can be seen the generating procedure can be very helpful in generating a varied set, as the objective filtering of the instances chooses a large number of non-literature function pairs that are demonstrably different.

An additional set of function pairs from the SOLAR simulation engine \citep{MScMLG} are also used in this study. This simulation engine simulates the behaviour of a solar power-plant, and provides truly black-box sources in the sense that sampling a source requires running processes that simulate the physical processes of the plant and for which no analytical expression is known. This simulator contains a variety of black-box objective functions, some of which are constrained and/or stochastic. The tenth objective function has 5 inputs, is constrained only to lie in a hypercube, is deterministic, and can be queried along with a fidelity value $fid \in (0,1]$, where setting $fid = 1$ returns the true value of the objective function. Defining $SOLAR_{fid}$ with $f \in \{0.1,0.2,\dots,0.8,0.9\}$ as the function pair $(f_h,f_l)$ where $f_h$ and $f_l$ query the objective function with fidelity values of 1 and $fid$ respectively, these additional nine function pairs are added to the 312 synthetic function pairs already chosen, bringing the total to 321 function pairs. Note that the SOLAR function pairs were not included in the instance filtering as they are all deemed to be valuable since they lead to ``real" black-box instances. Finding whether the synthetic instances are similar to the SOLAR instances can guide the creation of further synthetic instances.

Having chosen the set of function pairs, the set $\mathbf{I}$ is created using each of the 321 function pairs with $n_l \in \{4d, 8d, 12d, 16d, 20d\}$, $n_h \in \{2d, 4d, \dots, 18d, 20d\}$ and $n_h \leq n_l$, leading to a total of 9630 instances. As described in the previous section, for each instance 40 Kriging and Co-Kriging models are trained and their performances are compared using the Wilcoxon test. The remaining steps include choosing a set of relevant features, and the construction and analysis of the instance space. Before each of these steps, instance filtering is conducted to avoid bias both when choosing the features and when conducting the analysis as outlined by \citet{alipour2023enhanced}. Note that once again, certain instances are prioritised when choosing which instances to keep. Namely, the highest priority is given to SOLAR-based instances, the second-highest to literature-based instances, and the lowest priority is given to disturbance-based instances. The results are presented next.

\subsection{Algorithm performance analysis via ISA}

Using the automated feature selection procedure of the MATILDA toolkit leads to 9 features being chosen. This signifies that each instance lies in a 9-dimensional space, where each of the coordinates is the value of each of the 9 features. In order to visualise this space, a projection onto 2-dimensional space is achieved via the projection matrix given in Equation \ref{eq:projection}. Recall that the features $LCC^{0.2}_{sd}$, $LCC^{0.2^{1/d}}_{0.4}$, $LCC^{0.2^{1/d}}_{0.95}$ and $RRMSE$ all measure the quality of $f_l$ relative to $f_h$, and the feature $B^r$ is the ratio $\frac{n_h}{n_l}$, namely the amount of high-fidelity data available relative to the amount of low-fidelity data available. The remaining four features are landscape features, two of which characterise the landscape of $f_h$, whilst the other two characterise the landscape of the difference $f_h - f_l$ of the two sources. Further discussion on the features, their values and their impact on Kriging and Co-Kriging performance is given in Section \ref{sec:featureISA}. For now simply note that in all plots that follow, different properties of the constructed space in two dimensions are shown, where each point represents an instance and its coordinates $(z_1,z_2)$ represent only its projection in two dimensions of a 9-dimensional space.

\scriptsize
\begin{align}
\label{eq:projection}
\begin{bmatrix}
        z_1 \\
        z_2\\
    \end{bmatrix}
    =
    \begin{bmatrix}
       -0.4916 & -0.0889\\
        -0.3167 & -0.2321\\
        -0.1506 & 0.372\\
        -0.0568 & 0.4394\\
        0.1777 & -0.4154\\
        0.3696 & 0.0989\\
        0.4362 & 0.0526\\
        0.177 & 0.3381\\
        0.4031 & 0.2545\\
    \end{bmatrix}^\top
    \begin{bmatrix}
    B^r \\
    LCC^{0.2}_{sd} \\
    LCC^{0.2^{1/d}}_{0.4} \\
    LCC^{0.2^{1/d}}_{0.95}\\
    RRMSE\\
    f_h \text{, } MMCE^{0.5}_{lda}\\
    f_h \text{, } H_0\\
    f_h - f_l \text{, } \bar{R}^2_{L} \\
    f_h - f_l \text{, } \bar{R}^2_{LI} \\
    \end{bmatrix}
\end{align}

\normalsize
    
Figure \ref{fig:sources} shows the distributions of instances from different sources in the space. The benefit of supplementing literature instances with the disturbance-based procedure is clear in this plot, as despite not prioritising disturbance-based instances when applying the filtering procedure, many of them remain in the final set. This indicates that they are interesting either because they lie in regions of the space where traditional instances do not exist, or because they are similar to other instances but have different algorithm performance. Furthermore, it can be seen that both classical synthetic instances and disturbance-based instances can be similar to the truly black-box instances from the SOLAR simulator. However, the SOLAR instances lie towards the centre of the space, indicating that some synthetic instances that lie towards the border of the instance space might not represent realistic examples. This implies that whilst conclusions obtained from analysing algorithmic performance with synthetic instances will cover practical industrial cases, some extreme benchmarks might not be truly applicable to an applied setting.

\begin{figure}[!t]
    \centering
    \includegraphics[scale = 0.6]{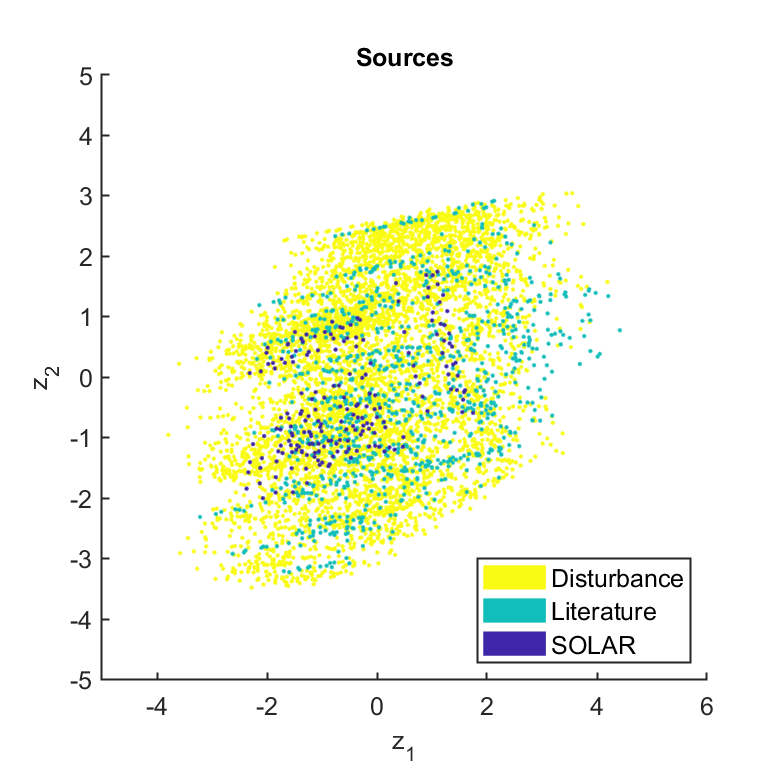}
    \caption{Sources of the function pairs used in each of the instances, where the dark blue points represent instances created from the SOLAR simulation, the light blue points represent classical literature instances, and the yellow points represent disturbance-based instances.}
    \label{fig:sources}
\end{figure}

\begin{figure}[!t]
    \centering
    \subfloat[]{
        \includegraphics[scale = 0.5]{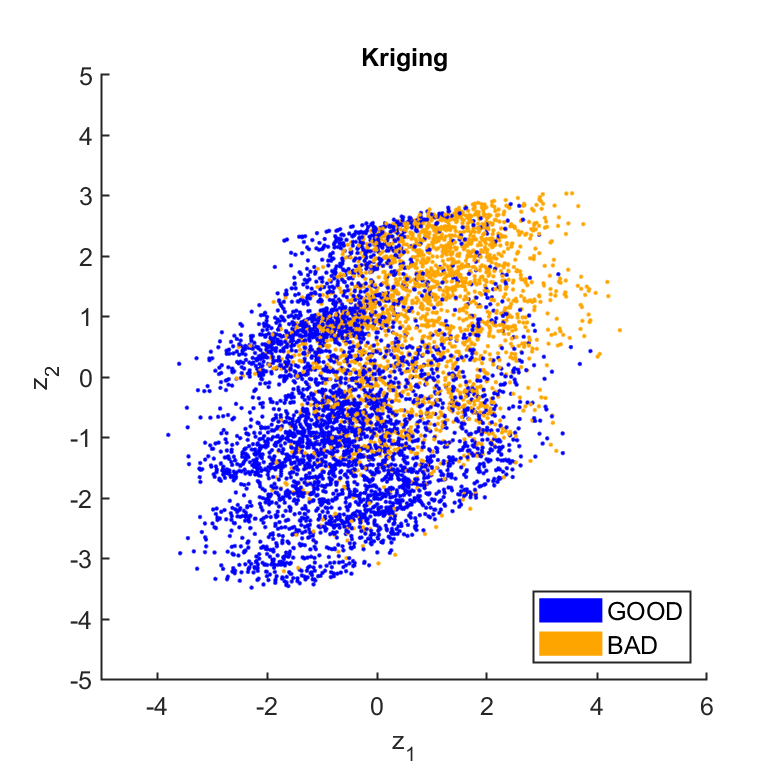}\label{fig:binaryPerformanceKriging}}
    \subfloat[]{
        \includegraphics[scale = 0.5]{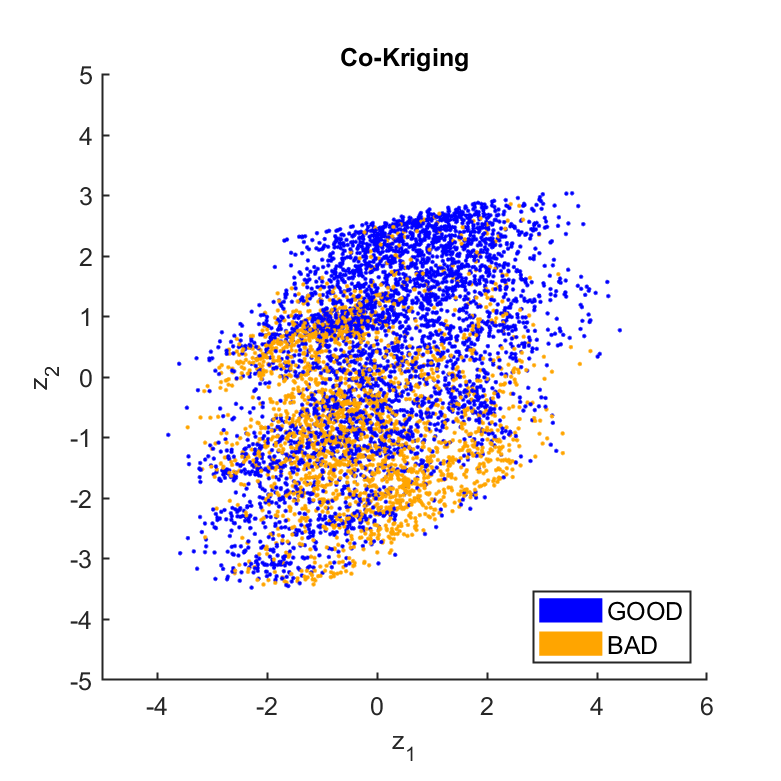}\label{fig:binaryPerformanceCoKriging}
    }
    \caption{Binary performance of \protect\subref{fig:binaryPerformanceKriging} Kriging and \protect\subref{fig:binaryPerformanceCoKriging} Co-Kriging models. The blue points represent instances for which the model's performance is labelled good, and the orange points represent instances for which the performance is labelled bad.}
\end{figure}

Figures \ref{fig:binaryPerformanceKriging} and \ref{fig:binaryPerformanceCoKriging} show the binary performance of the Kriging and Co-Kriging models respectively. Recall that a model is said to have good performance if it has a probability of at least 0.5 of having a higher or similar (within a tolerance of 0.001) correlation with $f_h$ than its competitor. Figure \ref{fig:binaryPerformanceKriging} clearly shows a top-right region where Kriging has bad performance, indicating that ignoring the low-fidelity data is counter-productive. On the other hand, Figure \ref{fig:binaryPerformanceCoKriging} shows that, whilst Co-Kriging performs well in the top part of the space, moving towards the bottom left leads to a region containing a lot of instances where Co-Kriging performs badly. It is worth stressing once again that it is possible for both models to perform well in the same instances, leading to blue regions in both plots. Therefore, despite the bottom left region containing instances where Co-Kriging performs well, Kriging performs well in almost all instances in the same region, indicating that the safer approach is to use Kriging and therefore ignore $f_l$. Regions that are predominantly blue in both figures indicate easier regions in the space in the sense that using either method is valid. The characteristics of these types of instances are discussed in Section \ref{sec:featureISA}.

\begin{figure}[!t]
    \centering
    \subfloat[]{
        \includegraphics[scale = 0.5]{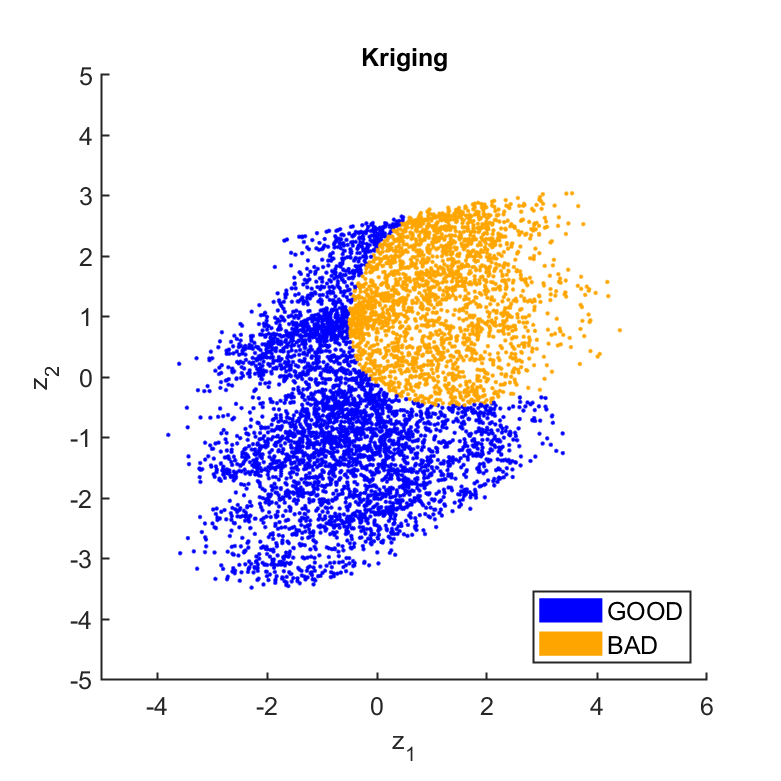}\label{fig:binaryPerformanceKrigingSVM}
    }
    \subfloat[]{
        \includegraphics[scale = 0.5]{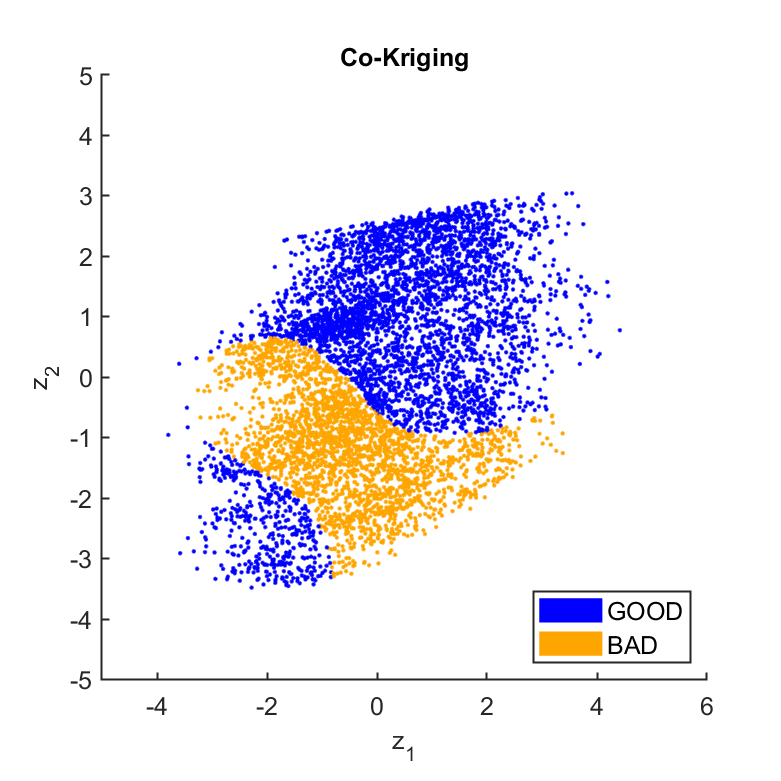}\label{fig:binaryPerformanceCoKrigingSVM}
    }
    \caption{SVM predictions of \protect\subref{fig:binaryPerformanceKrigingSVM} Kriging and \protect\subref{fig:binaryPerformanceCoKrigingSVM} Co-Kriging performance. Blue points represent instances where the model's performance is predicted to be good, and the orange points represent instances where the performance is predicted to be bad.}
\end{figure}



\begin{figure}[!t]
    \centering
    \includegraphics[scale = 0.6]{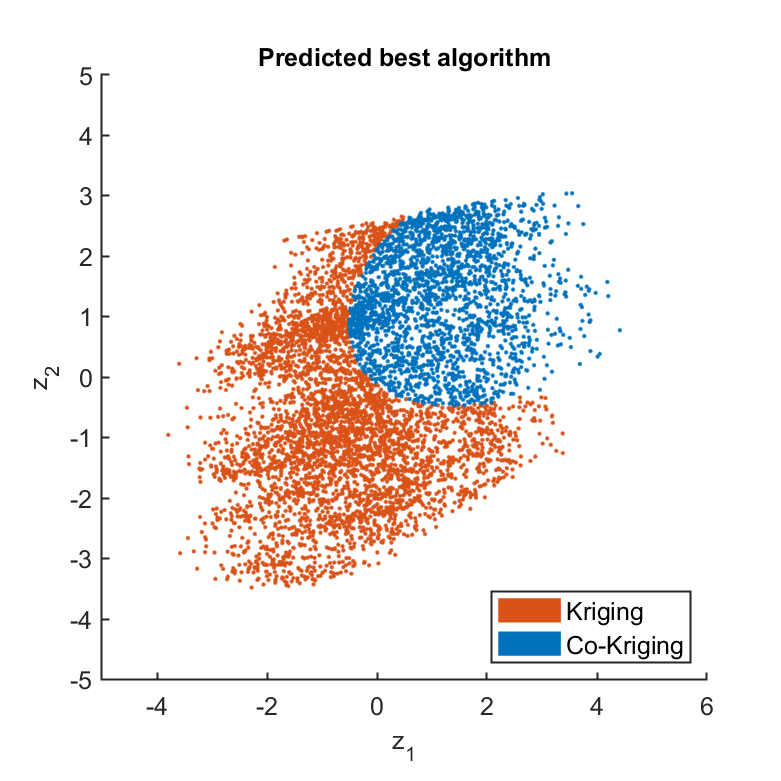}
    \caption{SVM selector which predicts whether Kriging or Co-Kriging should be used. The red dots indicate instances for which Kriging is predicted to be best, and the blue dots indicate instances for which Co-Kriging is predicted to be best.}
    \label{fig:predictorSVM}
\end{figure}

Given the good/bad labels of both techniques, Support Vector Machines (SVMs) are trained to predict when both Kriging and Co-Kriging can be used, as well as a selector for which method should be used for each of the regions of the space. The SVMs predictions for Kriging and Co-Kriging performance are shown in Figures \ref{fig:binaryPerformanceKrigingSVM} and \ref{fig:binaryPerformanceCoKrigingSVM}, respectively. It can be seen that in both the top-left and bottom-left regions both Kriging and Co-Kriging are predicted to be good, whereas only Co-Kriging is predicted to perform well in the top-right of the space and only Kriging is predicted to perform well in the bottom-right of the space. The best method predictions of the trained selector are shown in Figure \ref{fig:predictorSVM}. Perhaps unsurprisingly, the selector is almost identical to the SVM which predicts when Kriging performance is good or bad. Intuitively, this is the result of the division being much clearer for Kriging performance than for Co-Kriging performance. This is further reflected by Table \ref{tab:svms}, which presents the performance of all three SVMs. The constructed SVM which predicts Kriging performance has a much higher accuracy, precision and recall than that of the SVM which predicts Co-Kriging performance. In particular, a recall of $84.0\%$ indicates that the Kriging SVM is very accurate when predicting when Kriging will perform badly. As such, the selector chooses Co-Kriging when Kriging is predicted to perform poorly and Kriging the rest of the time. This implies it is a lot easier to know when a low-fidelity source will be beneficial than when it will be harmful, indicating that a low-fidelity source should only be used if one can assert it will be an asset.

\begin{table}[!t]
\centering
    \caption{Accuracy of the constructed SVMs which predict when each algorithm will perform well, as well as a selector which chooses an algorithm based on the location in the instance space. The second column shows the proportion of instances for which an algorithm is labelled good. The remaining columns show the statistical quality of the SVMs.}\label{tab:svms}%
    \begin{tabular}{@{}lllll@{}}
        \toprule
        Algorithm &  Pr(Good) & Accuracy & Precision & Recall\\
        \midrule
        Kriging & 0.610 & 77.4\% & 80.0\% & 84.0\%\\
        Co-Kriging & 0.618 & 70.2\% & 76.0\% & 75.6\%\\
        Selector & 0.806 & 80.6\% & - & -\\
        \toprule
    \end{tabular}
\end{table}

Finally, it is worth remarking that after filtering the metadata, the probability that Kriging and Co-Kriging will perform well is roughly the same i.e. 0.610 vs. 0.618, respectively. Therefore, simply choosing to ignore or always use the low-fidelity source is the correct strategy only slightly over $60\%$ of the time. The trained selector however correctly chooses a good model $80.6\%$ of the time, despite basing its decision only on the limited sample used to train the models. This accuracy is also remarkable due to the fact that it is derived from a 2-dimensional projection of a 9-dimensional space, as this is a simplification of the true characteristics of the instances. Furthermore, the use of the filtering procedure ensures the trained selector does not benefit from an (incorrectly) biased benchmark suite containing many instances that are similar to one another both in terms of feature values and algorithm performance. The constructed SVMs are therefore considered reliable enough to assess how different features affect algorithm performance in the next section.

\subsection{Analysis of features}
\label{sec:featureISA}

Before analysing the trends of the selected features against the predicted Kriging and Co-Kriging performance, informative conclusions can be drawn simply by looking at which features do not seem to help predict this performance. The automated feature selection of the implemented ISA toolkit follows a two-step process, first discarding any features that do not have an absolute correlation of at least 0.3 with either Kriging or Co-Kriging performance. Interestingly, all landscape features that characterise the low-fidelity source $f_l$ are discarded here, as the highest absolute feature correlation is 0.12, an extremely low value. Furthermore, the features $B_l$ and $B^r_l$ which indicate how much low-fidelity data is available have almost zero correlation with algorithm performance. The feature $B^r = \frac{n_h}{n_l}$ however has a correlation of 0.483 with Kriging performance, the highest correlation of all the features used. This indicates that when choosing whether a low-fidelity source is useful, the analysis should not focus on what this source looks like and how much data it provides. Rather, it should focus on how it behaves \textit{in relation to} the high-fidelity source $f_h$, and how much data it provides \textit{in relation to} the amount of high-fidelity data.

\begin{figure}[!t]
    \centering
    \includegraphics[scale = 0.5]{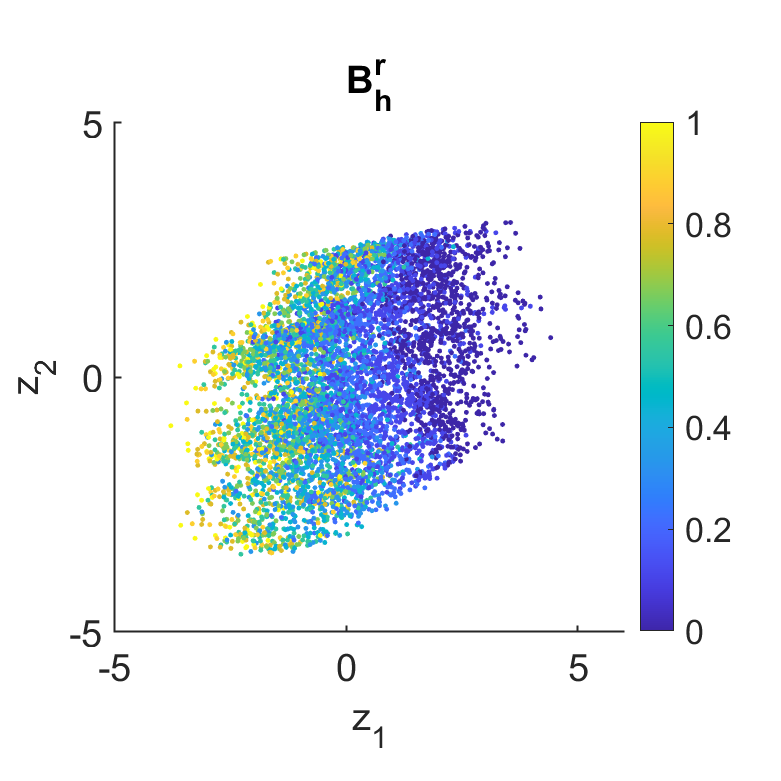}
    \caption{Feature distribution of the budget feature $B^r_h$. This feature indicates that going from the right to the left of the space means going through instances with the lowest to the highest relative amount of high-fidelity data. Note the colour grading is relative, with a value of 1 indicating a real feature value of 20, and a value of 0 indicating a real feature value of 2.}
    \label{fig:featHighBudget}
\end{figure}

The correlation of the remaining budget features $B_h = n_h$ and $B_h^r = \frac{n_h}{d}$ also provides a helpful insight. The feature $B_h$ measures how much high-fidelity data is available, but only has a correlation of 0.176 with algorithm performance. The feature $B_h^r$ however measures how much high-fidelity data is available relative to the problem dimension and has a correlation of 0.406, the second highest of all features being analysed. This indicates that studies on Bf-EBB problems should focus on sample budgets relative to problem dimension, something that intuitively makes sense and a large section of the literature is already doing. In fact, as shown in Figure \ref{fig:featHighBudget}, despite the feature $B_h^r$ not having been chosen to generate the instance space, analysing its value trend sheds light on the existence of regions on the left of the space where both Kriging and Co-Kriging performs well. These regions can be explained by the fact that when a lot of high-fidelity data is available, very accurate Kriging and Co-Kriging models can be trained and thus either technique is a good choice.

\begin{figure*}[!t]
    \centering
    \includegraphics[scale = 0.33]{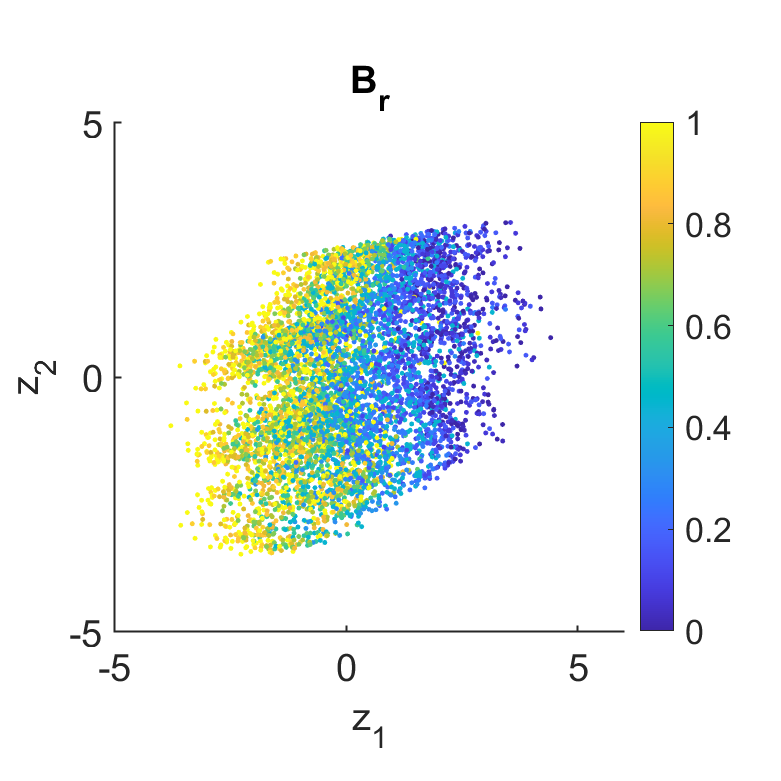}
    \includegraphics[scale = 0.33]{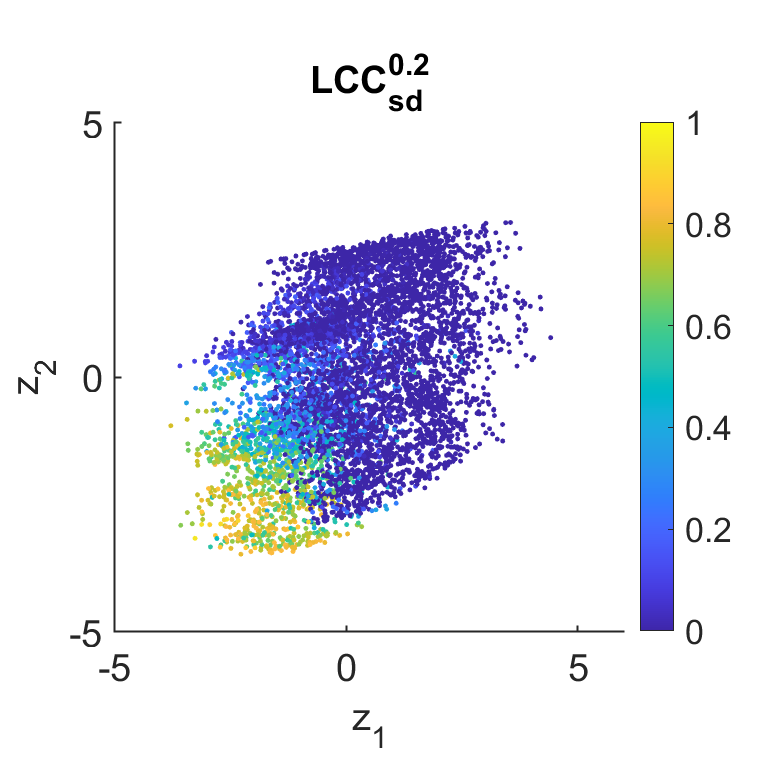}
    \includegraphics[scale = 0.33]{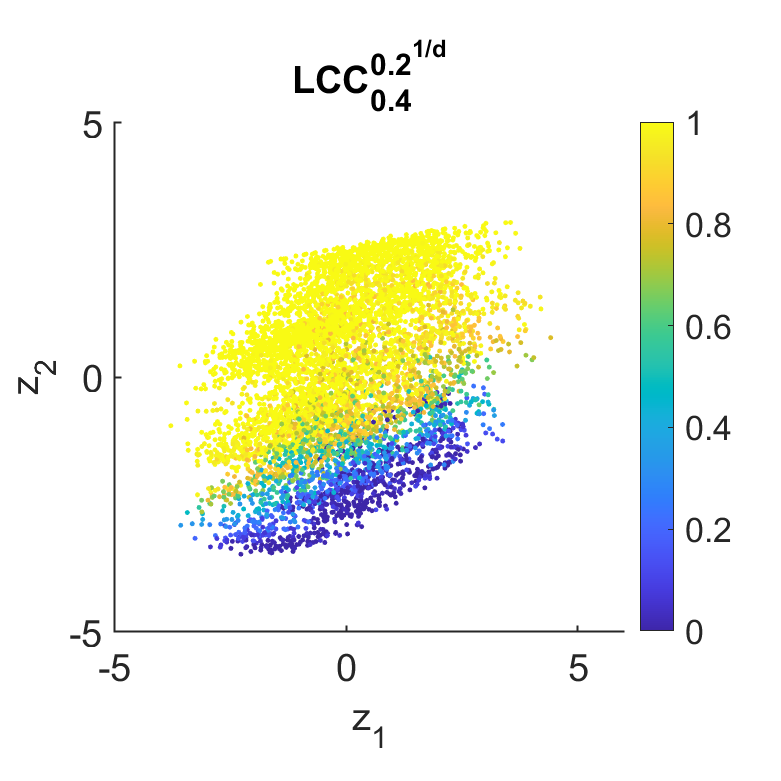}
    \includegraphics[scale = 0.33]{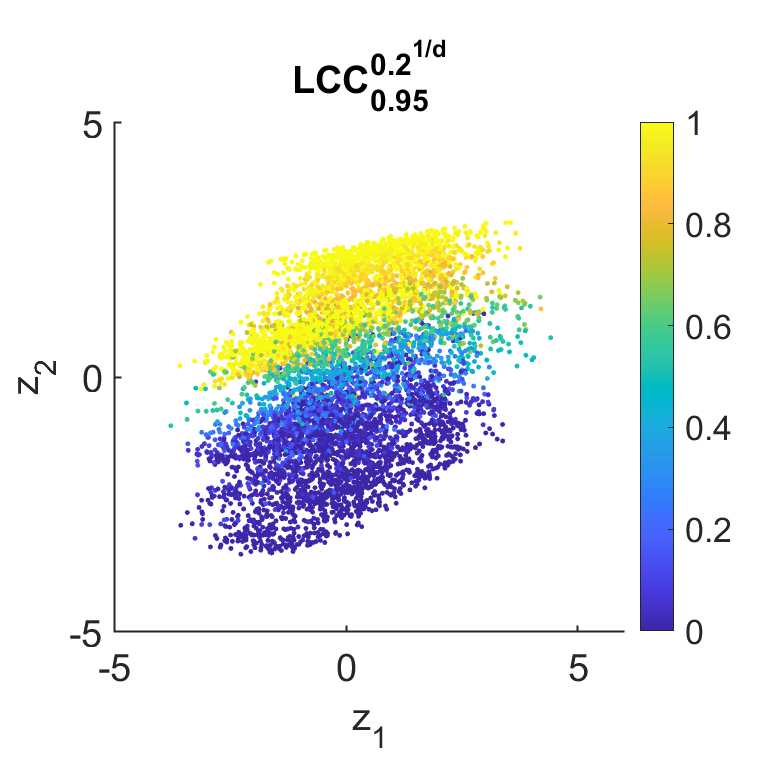}
    \includegraphics[scale = 0.33]{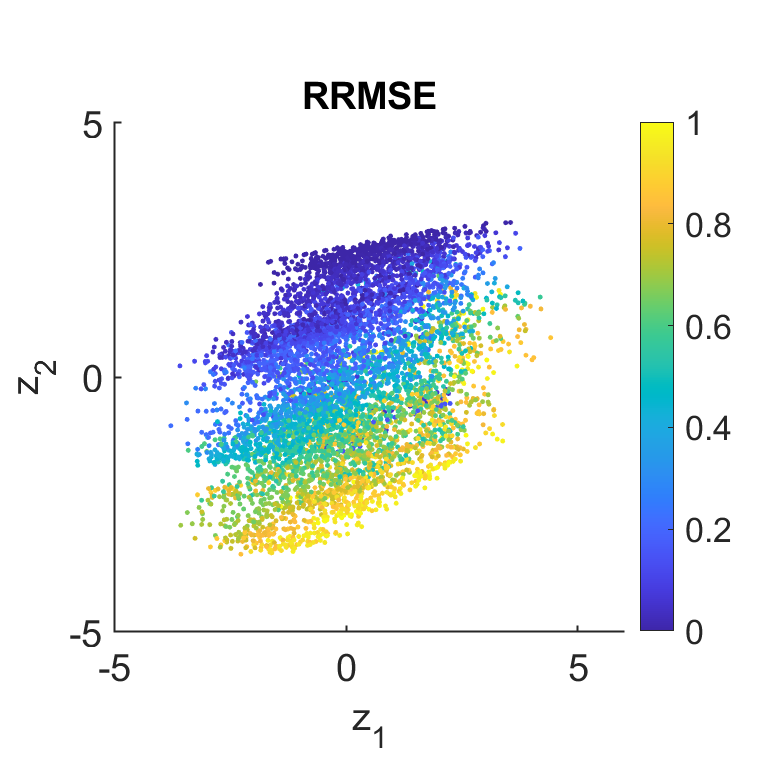}
    \includegraphics[scale = 0.33]{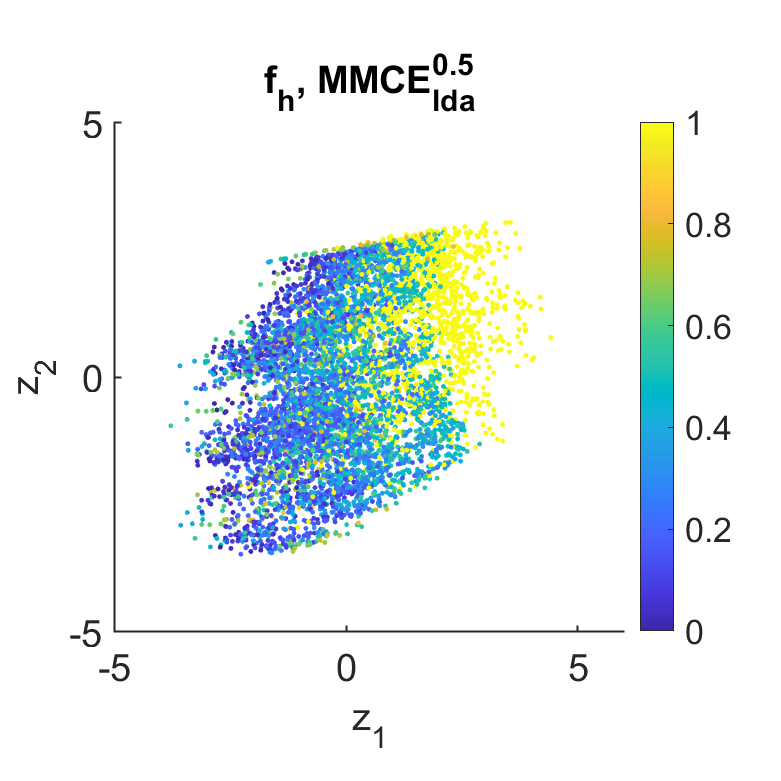}
    \includegraphics[scale = 0.33]{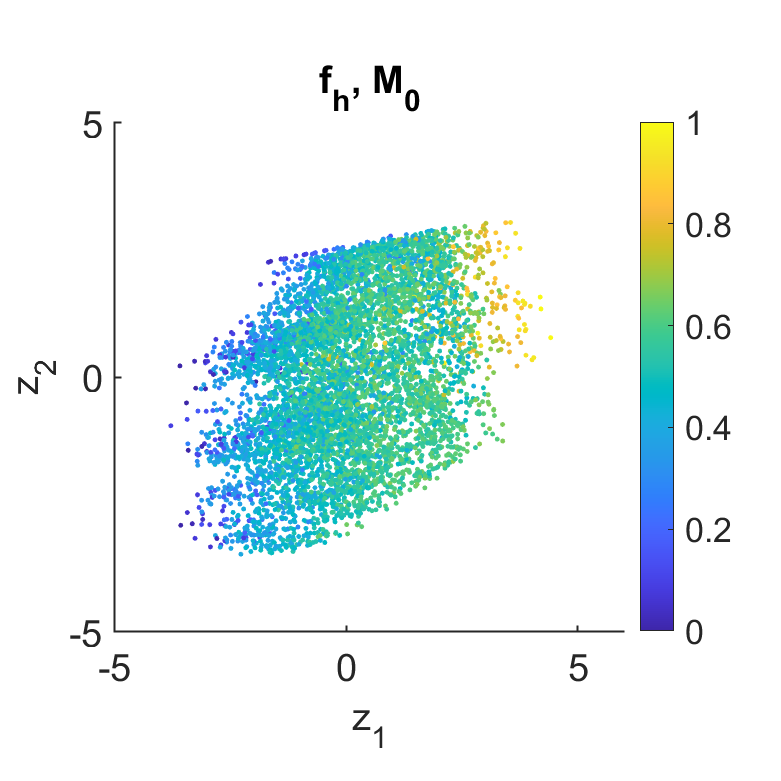}
    \includegraphics[scale = 0.33]{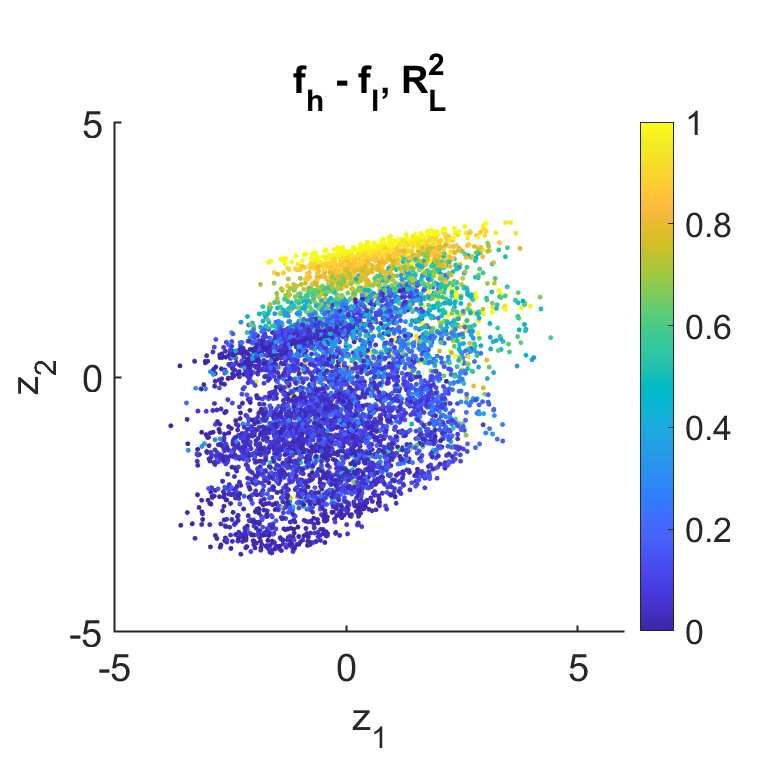}
    \includegraphics[scale = 0.33]{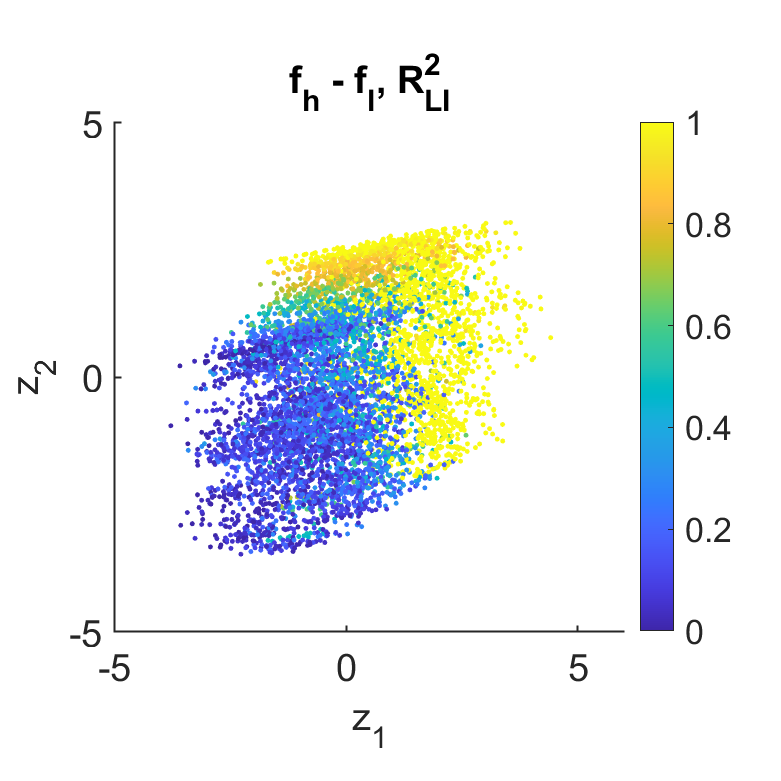}
    \caption{Feature distributions of the 9 chosen features. Note that the colour gradient is relative; for the range of each of the feature values consult Table \ref{tab:features}.}
    \label{fig:features}
\end{figure*}

Figure \ref{fig:features} illustrates the trends of the selected features within the space. Most features characterise the relationship between $f_h$ and $f_l$. Recall the $LCC^r$ features measure the local correlation characteristics of $f_l$ relative to $f_h$. The feature $LCC_{sd}^{0.2}$ measures the standard deviation of this local correlation using a relative radius of 0.2. The features $LCC^{0.2^{1/d}}_{0.4}$ and $LCC^{0.2^{1/d}}_{0.95}$ measure the probability that $f_l$ and $f_h$ have a local correlation of 0.4 and 0.95, respectively, using a relative radius of $0.2^{1/d}$. The feature $RRMSE$ measures the error between $f_h$ and $f_l$. The features $\bar{R}^2_{L}$ and $\bar{R}^2_{LI}$ are used to characterise the difference $f_h - f_l$ between sources, measuring how easy it is to model this difference using a simple linear model and a linear model with interactions, respectively. Finally, only two features characterise the landscape of $f_h$. The feature $MMCE_{lda}^{0.5}$ measures the error when separating the samples with the top and bottom $50\%$ $f_l$ values using a linear model. The feature $M_0$ measures the ruggedness of $f_h$. It is worth stressing here that the constructed space is by no means unique, especially since features other than these 9 passed the minimum correlation threshold of 0.3 with algorithm performance. Careful consideration of the created space however leads to good insights into the characterisation of a harmful low-fidelity source.

It appears overlaying the local correlation features $LCC^{0.2^{1/d}}_{0.4}$ and $LCC^{0.2^{1/d}}_{0.95}$ divides the space into a top, middle and bottom region. In the top region, $LCC^{0.2^{1/d}}_{0.95}$ values close to 1 indicate that $f_l$ has a very high probability of being very accurate locally (i.e. having a correlation of at least 0.95 locally). This corresponds to a region where Co-Kriging can be used, and therefore the low-fidelity source is valuable. In the bottom region, $LCC^{0.2^{1/d}}_{0.4}$ values close to 0 indicate that $f_l$ has a very low probability of being somewhat accurate locally (i.e. having a correlation of at least 0.4 locally). This corresponds to a region where Co-Kriging should not be used, and therefore the low-fidelity source is harmful. The middle region corresponds to instances that are often somewhat locally accurate, but rarely very locally accurate. For these instances, Co-Kriging should only be used if little high-fidelity data is available. It is worth noting that only $LCC^r$ features calculated with a relative radius of $0.2^{1/d}$  seem to help predict algorithm performance. Despite the recommendation by \citet{andres2022bifidelity} to set $r = 0.2$, in the applied setting using this relative radius leads to small neighbourhoods which contain very few sample points, and lead to unchanging local correlation values. The $LCC^{0.2}_{sd}$ plot in Figure \ref{fig:features} illustrates this, as for most of the space the calculated standard deviation of the local correlation is 0. Therefore, in an applied setting a relative radius of $0.2^{1/d}$ which grows with the problem dimension should be used when calculating these features.

The pair of features $\bar{R}^2_{L}$ and $\bar{R}^2_{LI}$ are also helpful in predicting when a low-fidelity source can be used. When the value of these features is high, this indicates the difference between the two sources is easy to model. In these cases, Co-Kriging should be used. Intuitively this makes sense, as training a Co-Kriging model can be seen as the combination of training a model of $f_l$, and training a model of the difference $f_h-f_l$. Therefore, if most of the complex behaviour of $f_h$ is represented by $f_l$ (and therefore $f_h - f_l$ is simple), $f_l$ can be very beneficial. It is important to note that perhaps the feature $\bar{R}^2_{L}$ should be relied upon rather than the feature $\bar{R}^2_{LI}$, as for very small sample sizes a linear model with interactions can fit the sample perfectly. This leads to cases in the bottom right of the space where the $\bar{R}^2_{LI}$ value is 1, but Co-Kriging performs worse than Kriging. Finally, note that a low $\bar{R}^2_{L}$ or $\bar{R}^2_{LI}$ value does not necessarily mean Co-Kriging cannot be used.

Not all selected features are necessarily helpful. The landscape features $MMCE_{lda}^{0.5}$ and $M_0$, in particular, appear to be strongly affected by the sample size used to calculate them, as their trend is very similar to the trend of the budget feature $B_h^r$. It is therefore not recommended to use these two features when predicting Kriging and Co-Kriging performance. Furthermore, despite the error feature $RRMSE$ seemingly giving a nice separation of the space, this is the only selected feature to be unbounded and therefore the only selected feature to have been normalised. It is hard to predict the impact of this feature value for instances with larger $RRMSE$ than the benchmark suite being analysed. As such, it is also not recommended to use this feature when predicting algorithm performance. \\

\subsection{Proposal of new guidelines}

The trained SVMs of the previous subsection provide predictions of regions where Kriging and Co-Kriging should be used based on the 2-dimensional projection. This has the advantage of being intuitive for users to identify the driving features in algorithm performance. It can also be beneficial however to analyse how single feature values impact model performance in order to derive some simple rules that industrial practitioners can follow.

\begin{figure}[!t]
    \centering
    \includegraphics[scale = 0.6]{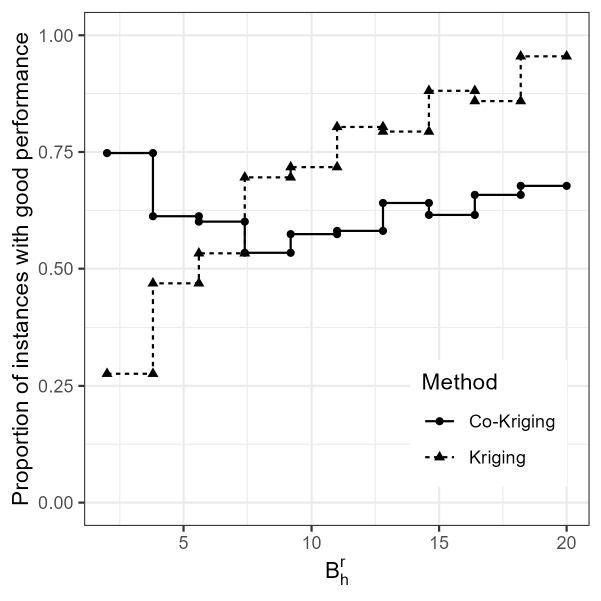}
    \caption{Proportion of instances for which Kriging and Co-Kriging perform well for different $B^r_h$ values, the amount of high-fidelity data relative to problem dimension.}
    \label{fig:proportionBh}
\end{figure}

\begin{figure}[!t]
    \centering
    \includegraphics[scale = 0.6]{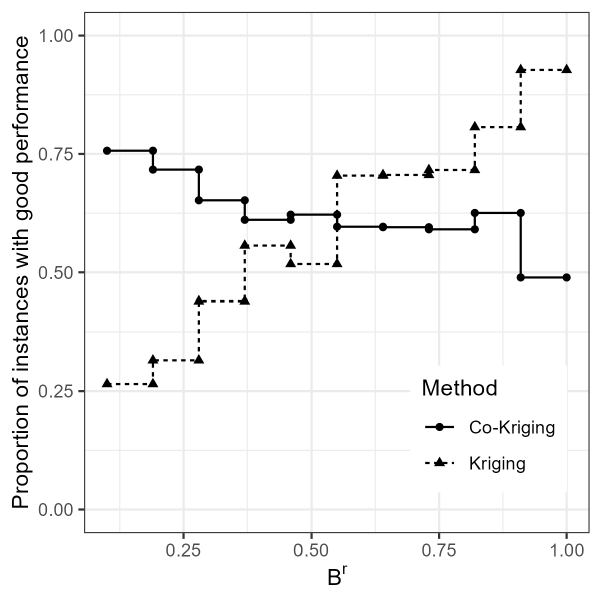}
    \caption{Proportion of instances for which Kriging and Co-Kriging perform well for different $B^r$ values, the amount of high-fidelity data available relative to low-fidelity data available.}
    \label{fig:proportionBr}
\end{figure}

In order to generate these guidelines, the proportion of instances with good Kriging/Co-Kriging performance is plotted against a variety of feature value bins. In each of the plots, the lowest and highest $x$-axis value of a flat segment represents the range of a bin, and the $y$-axis value represents the proportion of instances within that range which are labelled as good for a particular model. The first such graph is shown in Figure \ref{fig:proportionBh}. This graph indicates that the amount of high-fidelity data available alone is not a very strong indicator of whether Co-Kriging should be used. It does however show that when a lot of high-fidelity data is available, it is almost guaranteed that the best decision is to not consider other data sources when constructing a model. When the relative amount of high-fidelity data available is between 18 and 20 times the dimension of the problem, Kriging is a good choice for $95\%$ of the instances. Further analysis of harmful data sources should therefore be restricted to cases where at most $20d$ high-fidelity samples are available. Figure \ref{fig:proportionBr} indicates that in order to rely on a low-fidelity source, it is important to have a lot more low-fidelity data than high-fidelity data. In particular, training a Co-Kriging model when $n_l = n_h$ is to be avoided, as in this special case Kriging has good performance for $93\%$ of the instances. Based on this graph, it is recommended to have at least 2.5 times as much low-fidelity as high-fidelity data when using Co-Kriging.

\begin{figure}[!t]
    \centering
    \includegraphics[scale = 0.6]{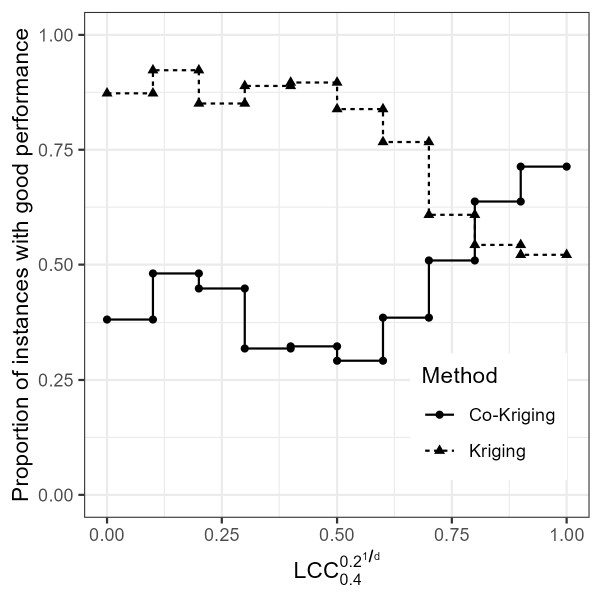}
    \caption{Proportion of instances for which Kriging and Co-Kriging perform well for different $LCC^{0.2^{1/d}}_{0.4}$ values.}
    \label{fig:proportionLCC4}
\end{figure}

\begin{figure}[!t]
    \centering
    \includegraphics[scale = 0.6]{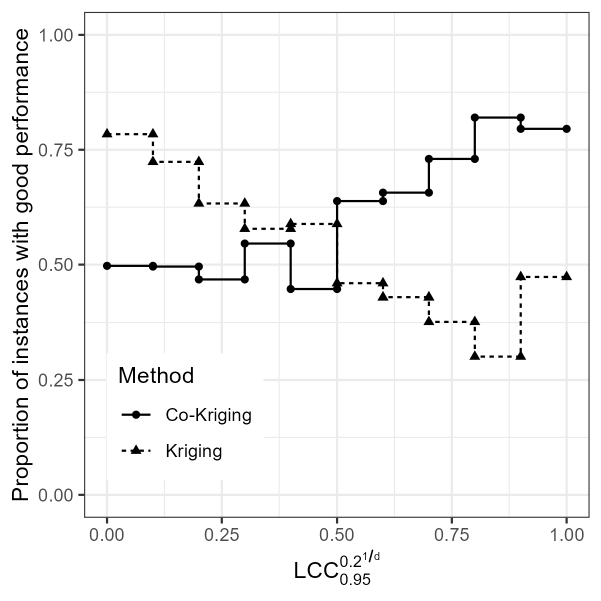}
    \caption{Proportion of instances for which Kriging and Co-Kriging perform well for different $LCC^{0.2^{1/d}}_{0.95}$ values.}
    \label{fig:proportionLCC95}
\end{figure}

\begin{figure}[!t]
    \centering
    \includegraphics[scale = 0.6]{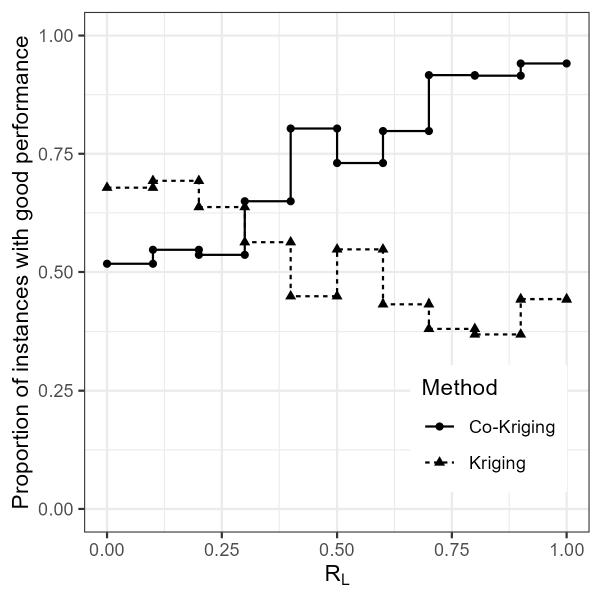}
    \caption{Proportion of instances for which Kriging and Co-Kriging perform well for different $\bar{R}^2_{L}$}
    \label{fig:proportionr2L}
\end{figure}

Figures \ref{fig:proportionLCC4}, \ref{fig:proportionLCC95} and \ref{fig:proportionr2L} provide good cut-off points for three feature values when choosing which model to use. In particular, Figure \ref{fig:proportionLCC4} shows that when $LCC^{0.2^{1/d}}_{0.4} \leq 0.7$, Kriging performs well for at least $76\%$ of the instances in each of the feature bins. Similarly, Figure \ref{fig:proportionLCC95} shows that when $LCC^{0.2^{1/d}}_{0.95} \geq 0.7$, Co-Kriging performs well for at least $73\%$ of the instances in each of the feature bins. Finally, Figure \ref{fig:proportionr2L} shows that when $\bar{R}^2_{L} \geq 0.4$, Co-Kriging also performs well for at least $73\%$ of the instances in each of the feature bins.

\begin{figure}[!t]
    \centering
    \includegraphics[scale = 0.6]{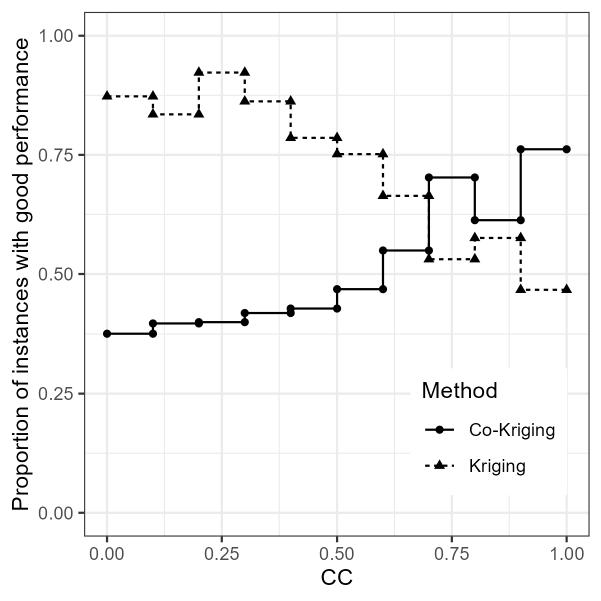}
    \caption{Proportion of instances for which Kriging and Co-Kriging perform well for different $CC$ values, the global correlation between $f_l$ and $f_h$.}
    \label{fig:proportionCC}
\end{figure}

Before combining these insights into simple rules to decide which model to use, it is worth also looking at the feature $CC$. This feature measures the overall correlation between $f_h$ and $f_l$, giving a sense of the global quality of $f_l$. It was shown by \citet{toal2015some} to provide a good indication of when $f_l$ can be relied upon, and is often used in the literature when assessing low fidelity sources. Despite not having been chosen to generate the instance space, Figure \ref{fig:proportionCC} shows that it is better to construct Co-Kriging models when $CC \geq 0.7$, although doing so only when $CC \geq 0.9$ as Toal recommends is a more conservative estimate. Relying on the basic rule of constructing Co-Kriging models when $CC \geq 0.7$ and Kriging models otherwise leads to choosing the right model $75.0\%$ of the time. Whilst this accuracy is not as large as that of the selector trained in the previous subsection, it is a marked improvement over always choosing the same model regardless of instance characteristics. This can be considered the performance of current literature guidelines.

Moving now to the proposal of new basic guidelines based on the insights of this section, the following rules are proposed when choosing whether to construct a Kriging or a Co-Kriging model:

\begin{enumerate}
    \item For instances where either $B^r_h \geq 18$, $B^r = 1$ or $LCC^{0.2^{1/d}}_{0.4} \leq 0.7$, Kriging should be used.
    \item For instances which do not satisfy the above conditions, but for which $LCC^{0.2^{1/d}}_{0.95} \geq 0.5$, or $\bar{R}^2_{L} \geq 0.4$, Co-Kriging should be used.
    \item For instances which are not covered by the previous two sets of rules, Co-Kriging should be used when $B^r_h \leq 5$, and Kriging should be used otherwise.
\end{enumerate}

These three sets of rules can be reduced to the following: Kriging should be used when plenty of data is available, the same amount of $f_l$ and $f_h$ data is available, or if $f_l$ appears to be clearly harmful. If this is not the case, and $f_l$ appears to be clearly beneficial, Co-Kriging should be used. In the remainder of the cases, that is when $f_l$ is somewhat helpful, Co-Kriging should be used only if limited high-fidelity data is available.

\begin{figure}[!t]
    \centering
    \includegraphics[scale = 0.6]{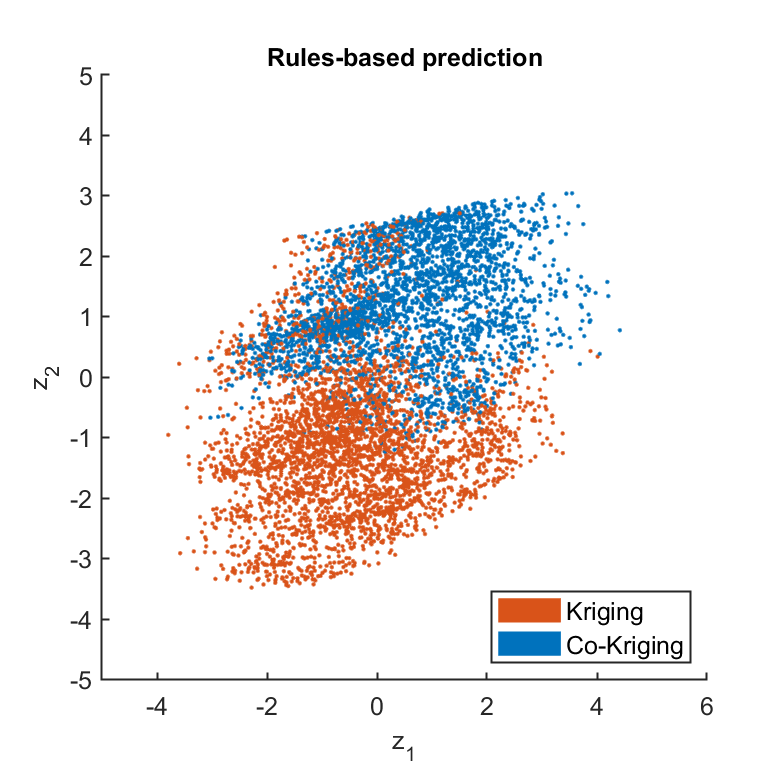}
    \caption{Selector which predicts whether Kriging or Co-Kriging should be used based on the simple rules given in this section. The red dots indicate instances for which Kriging is predicted to be best, and the blue dots indicate instances for which Co-Kriging is predicted to be best. Note that contrary to the predictor shown in Figure \ref{fig:predictorSVM}, no clear regions arise from this prediction.}
    \label{fig:predictorSimpleRules}
\end{figure}

Applying these guidelines to the filtered benchmark set containing 6500 instances yields very high accuracies. In particular, 2477 of those instances fall in the category defined by the first set of rules. For these instances, Kriging is recommended and is indeed labelled as having good performance $87.8\%$ of the time. A total of 2503 instances fall in the category defined by the second set of rules. For this second set of instances, Co-Kriging is recommended, and it is labelled as a good algorithm $81.6\%$ of the time. This indicates that the first two sets of rules not only cover a very large set of instances, but also provide a recommendation which is highly accurate. The accuracy of applying the full set of rules to the entire filtered benchmark set leads to a good algorithm being chosen $81.4\%$ of the time, which is an improvement over the already highly accurate selector of the previous subsection. To see why this is the case, it is worth plotting the algorithm prediction within the space, which is shown in Figure \ref{fig:predictorSimpleRules}. It can be seen here that following these sets of rules leads to a prediction which contains a fair amount of ``confetti", something the selector trained in the previous section is not allowed to do. The reason behind this is that despite the rules leading to a higher accuracy, they do not provide the powerful intuitive insights inherent to ISA. When applying Kriging and Co-Kriging to industrial problems however, these sets of rules can provide an accurate suggestion of the best model to use based on the limited information available.

\section{Conclusion}

This study has characterised beneficial and harmful low-fidelity sources when modelling an expensive black-box $f_h$. Previous studies that have conducted similar work have done so under perfect conditions, in the sense that the analysis conducted relied on a large amount of data that is unavailable in practice. In this study, this characterisation is constructed using only the limited information available to train a model, meaning both the generated simple guidelines and the more intricate instance space can be directly applied to industrial problems. Despite relying only on limited information, however, the trained predictor still achieves an $80.6\%$ accuracy when predicting which model will be most accurate. This is a marked improvement over choosing to always use either Kriging or Co-Kriging, as both options perform well only about $61\%$ of the time. This is particularly remarkable as the predictor is trained on a 2-dimensional projection of a 9-dimensional space that, despite leading to more intuitive predictions, is often at the cost of a loss in prediction accuracy. Using the insights of ISA and departing from the restriction of this 2-dimensional projection, a set of simple guidelines have been developed which show a further improvement on model prediction. Despite being based on easy to calculate features approximated with limited data, these rules lead to a good model being chosen $81.6\%$ of the time, a big improvement over existing guidelines.

The characterisation of a harmful source has been achieved by comparing the performance of (single-source) Kriging models with (two-source) Co-Kriging models. The widespread usage of these models and the fact that many new techniques are either an extension or rely on a similar framework implies the findings of this work are very likely applicable to other models in the literature. It is important to note however that analysis of ISA is iterative in nature. Further instances and features can be developed, as well as the analysis extended to other models, to further the understanding of how low-fidelity sources can be exploited.

Finally, the work presented here has focused on the simplest variant of Bf-EBB problems, namely the case where the data has already been gathered. In this case, the only decision available to the practitioner is which model to train. In many cases, a total budget is supplied instead. Further research will shift its attention to this more complex scenario, where the user needs to decide how to split the budget between sources, and where to gather further samples in the space. Future work will use the findings of this work to develop adaptive techniques that dynamically choose when to rely on low-fidelity sources. The development of these techniques will focus both on optimisation, and on the case where the sole aim is to construct as accurate a model as possible. We believe the findings of this work will allow newly developed techniques to outperform existing methods that choose to always or never use a source.

\section*{Statements and declarations}

\textbf{Funding:} This research was supported by the Australian Research Council under grant number IC200100009 for the ARC Training Centre in Optimisation Technologies, Integrated Methodologies and Applications (OPTIMA). The first author is also supported by a Research Training Program Scholarship from The University of Melbourne.

\textbf{Acknowledgements:} This research was supported by The University of Melbourne’s Research Computing Services and the Petascale Campus Initiative.

\textbf{Conflict of interest:} The authors declare that they have no conflict of interest.

\textbf{Replication of results:} The code implemented for this study is divided in two modular repositories. The first repository \citep{nandres2022benchmarks} implements all of the referenced function pairs, provides the code used to measure the feature values, and generates the benchmark suite with 321 instances, as well as additional benchmark suites of sizes 65, 100, 140, 206 and 509. Researchers wanting to assess the performance of other techniques on these benchmarks are directed to this repository. The second repository \citep{nandres2022methods} implements both Kriging and Co-Kriging, and the framework to analyse their performance on the selected benchmark suite. Researchers wanting to rerun these experiments, or to use the implementation of Kriging and Co-Kriging (as well as other techniques) are directed to this repository. All relevant metadata (i.e. features and algorithm performance) is also made available in the relevant repositories for further analysis if desired.

\appendix

\section{Feature definitions}\label{secA1}

The work of \citet{toal2015some} proposes two features which measure the global quality of $f_l$ relative to $f_h$, namely the Correlation Coefficient $(CC)$ and the Root Mean Squared Error $(RMSE)$. Both features are calculated on a given set of sample points $\mathbf{X} = \{\mathbf{x}_1,\dots,\mathbf{x}_n\} \subset \Omega$. The objective function values of both $f_h$ and $f_l$ are used to calculate these features, namely the sets $\mathbf{y}_l = \{f_l(\mathbf{x}_1),\dots,f_l(\mathbf{x}_n)\}$ and $\mathbf{y}_h = \{f_h(\mathbf{x}_1),\dots,f_h(\mathbf{x}_n)\}$. The definition of both features is the following

\begin{align*}
    &RMSE = \left[\frac{\sum_{i=1}^n (f_l(\mathbf{x}_i) - f_h(\mathbf{x}_i))^2}{n} \right]^{1/2}\\
    CC &= \left[\frac{1}{n-1}\left(\frac{\sum_{i = 1}^{n} (f_l(\mathbf{x}_i) - \bar{y}_l)(f_h(\mathbf{x}_i) - \bar{y}_h)}{s_{y_l} s_{y_h}}\right)\right]^2\\[1ex]
    \bar{y}_l &= \frac{1}{n}\sum_{i = 1}^{n}f_l(\mathbf{x}_i)\\
    s_{y_l} &= \left[\frac{\sum_{i = 1}^{n}(f_l(\mathbf{x}_i) - \bar{y}_l)^2}{n-1}\right]^{1/2}\\
    \bar{y}_h &= \frac{1}{n}\sum_{i = 1}^{n}f_h(\mathbf{x}_i)\\
    s_{y_h} &= \left[\frac{\sum_{i = 1}^{n}(f_h(\mathbf{x}_i) - \bar{y}_h)^2}{n-1}\right]^{1/2}\\
\end{align*}

A high $CC$ value indicates that overall, $f_l$ behaves similarly to $f_h$. A low $RMSE$ value indicates that overall, the two sources differ little in the space. The work of \citet{andres2022bifidelity} proposes a scaling of the $RMSE$ feature in order to make it comparable between pairs of functions. This scaled feature is denoted Relative $RMSE$ ($RRMSE$) and is given by 

\begin{align*}
    RRMSE = \frac{RMSE}{\max\{\mathbf{y}_h\} - \min\{\mathbf{y}_h\}}\\
\end{align*}

The same authors propose a set of features which assess the local correlation characteristics of $f_l$ and $f_h$. First the definition of Weighted Correlation Coefficient ($WCC(\mathbf{w}$)) is given for a set of weights $\mathbf{w} = \{w_1,\dots,w_n\}$

\begin{align*}
    WCC(\mathbf{w}) &= \left[\frac{1}{\sum_{i=1}^{n} w_i}\left(\frac{S}{s_{y_l} s_{y_h}}\right)\right]^2\\
    S &= \sum_{i = 1}^{n} w_i(f_l(\mathbf{x}_i) - \bar{y}_l)(f_h(\mathbf{x}_i) - \bar{y}_h)\\
    \bar{y}_l &= \frac{\sum_{i=1}^{n} w_i f_l(\mathbf{x}_i)}{\sum_{i=1}^{n} w_i}\\
    s_{y_l} &= \left[\frac{\sum_{i=1}^{n} w_i (f_l(\mathbf{x}_i) - \bar{y}_l)^2}{\sum_{i=1}^{n} w_i}\right]^{1/2}\\
    \bar{y}_h &= \frac{\sum_{i=1}^{n} w_i f_h(\mathbf{x}_i)}{\sum_{i=1}^{n} w_i}\\
    s_{y_h} &= \left[\frac{\sum_{i=1}^{n} w_i (f_h(\mathbf{x}_i) - \bar{y}_h)^2}{\sum_{i=1}^{n} w_i}\right]^{1/2}\\
\end{align*}

Next the Local Correlation Coefficient at a point $\mathbf{x}$ with radius $r$ is defined as

\begin{align*}
    LCC^r(\mathbf{x}) &= WCC(\mathbf{w})\\
    \text{where} \qquad w_i &= \min\{0, 1-\frac{\|\mathbf{x} - \mathbf{x}_i\|}{r\|\mathbf{x}^\top - \mathbf{x}^\bot\|}\}\\
\end{align*}

It is recommended to scale the data to lie inside the unit hypercube $[0,1]^d$ before calculating this feature. This measure calculates the correlation between $f_l$ and $f_h$ inside the $d$-sphere centred at a point $\mathbf{x}$ with radius $r\|\mathbf{x}^\top - \mathbf{zx}^\bot\|$. The weights lead to a higher impact of the correlation of points closer to the center. Now define the sets $\mathcal{L}^r = \{LCC^r(\mathbf{x}_1), \dots, LCC^r(\mathbf{x}_n)\}$ and $\mathcal{L}^r_p = \{LCC^r(\mathbf{x}) \in \mathcal{L}^r \quad | \quad LCC^r(\mathbf{x}) \geq p\}$, which are used to define the features

\begin{align*}
    LCC^r_p &= \frac{|\mathcal{L}^r_p|}{|\mathcal{L}^r|}\\
    LCC^r_{mean} &= \frac{1}{n}\sum_{i = 1}^n LCC^r(\mathbf{x}_i)\\
    LCC^r_{sd} &= \sqrt{\frac{\sum_{i=1}^{n} \left[LCC^r(\mathbf{x}_i) - LCC^r_{mean}\right]^2}{n-1}}\\
    LCC^r_{coeff} &= \frac{LCC^r_{sd}}{LCC^r_{mean}}\\    
\end{align*}

The $LCC^r_p$ calculate the probability that $f_l$ has a local correlation of at least $p$ with $f_h$, whereas the features $LCC^r_{mean}$, $LCC^r_{sd}$ and $LCC^r_{coeff}$ calculate various distribution measures of this local correlation.


\bibliographystyle{plainnat}  
\bibliography{bibliography.bib}

\end{document}